\documentclass[prx,twocolumn,groupedaddress,showpacs]{revtex4}

\usepackage{graphicx,amsmath}
\usepackage{subfigure}
\usepackage{epstopdf}
\usepackage{hyperref}

\usepackage{color}

\begin{document}

\title{The Phase Diagram of the $\nu=5/2$ Fractional Quantum Hall Effect: Effects of Landau Level Mixing
and Non-Zero Width}

\author{Kiryl Pakrouski$^{1}$, Michael R. Peterson$^{2}$, Thierry Jolicoeur$^{3}$, Vito W. Scarola$^{4}$, Chetan Nayak$^{5,6}$ and Matthias Troyer$^{1}$}
\affiliation{$^{1}$Theoretische Physik, ETH Zurich, 8093 Zurich, Switzerland}
\affiliation{$^{2}$Department of Physics \& Astronomy, California State University Long Beach,  Long Beach, California 90840, USA}
\affiliation{$^{3}$ Laboratoire de Physique Th\'eorique et Mod\`eles Statistiques,
CNRS and Universit\'e Paris-Sud, 91405 Orsay, France}
\affiliation{$^{4}$Department of Physics, Virginia Tech, Blacksburg, Virginia 24061, USA}
\affiliation{$^{5}$Department of Physics, University of California, Santa Barbara, California 93106, USA}
\affiliation{$^{6}$Microsoft Research, Station Q, Elings Hall, University of California, Santa Barbara, California 93106, USA}

\begin{abstract}
Interesting non-Abelian states, e.g., the Moore-Read Pfaffian and the anti-Pfaffian, offer candidate descriptions of the $\nu = 5/2$ 
fractional quantum Hall state.  But the significant controversy surrounding the nature of the $\nu = 5/2$ state has been hampered by the fact that the competition between these and other states is affected by small parameter changes. To 
study the phase diagram of the $\nu = 5/2$ state we numerically diagonalize a comprehensive effective Hamiltonian
describing the fractional quantum Hall effect (FQHE) of electrons under realistic conditions in GaAs semiconductors.  The
effective Hamiltonian takes Landau level mixing into account to lowest-order perturbatively in $\kappa$, the ratio of the Coulomb energy scale to the cyclotron gap.  We also incorporate non-zero width $w$ of the quantum well and sub-band mixing.  We find the ground state in both the torus and spherical geometries as a function of $\kappa$ and $w$.  To sort out the non-trivial competition between candidate ground states we analyze the following 4 criteria:
its overlap with trial wave functions; the magnitude of energy gaps;
the sign of the expectation value of an order parameter for particle-hole symmetry breaking; and the entanglement spectrum.  We conclude that the ground state is in the universality class of the Moore-Read Pfaffian state, rather than the anti-Pfaffian, for $\kappa < {\kappa_c}(w)$, where ${\kappa_c}(w)$ is a $w$-dependent critical value $0.6 \lesssim{\kappa_c}(w)\lesssim 1$.  We observe that both Landau level mixing
and non-zero width suppress the excitation gap, but Landau level mixing has a larger effect in this regard.  Our findings have important implications for the identification of non-Abelian fractional quantum Hall states.
\end{abstract}

\date{\today}


\maketitle

\section{Introduction}
The $\nu=5/2$ fractional quantum Hall state is well-established: it has a robust energy gap and has been
observed in a large number of different GaAs samples \cite{Willett87,Willett88,Eisentstein88,Eisenstein90,Pan99,Pan01,Eisenstein02,Xia04,Choi08,Pan08,Radu08,Dean08,Dolev08,Kumar10,Nuebler10,Pan11,Liu11,Nuebler-PRL2012,Liu13,Gamez13,Pan2014,Deng14,Reichel14}, yet its underlying quantum order remains
 mysterious.  Although there is strong evidence that the ground state is spin-polarized \cite{Tiemann11} with a fractional quasiparticle effective charge of $e/4$ \cite{Radu08,Dolev08,Willett09,Venkatachalam11}, there are some experiments that remain difficult to interpret in
this light \cite{Stern10,Rhone10,Baer2014}.  Perhaps the most interesting hypothesized
property of this state -- non-Abelian quasiparticle braiding \cite{Nayak96c,Read96,Tserkovnyak03,Seidel08,Read08,Baraban09,Prodan09,Bonderson11a} -- is controversial.
There are experiments consistent with non-Abelian quasiparticles \cite{Willett09,Willett10,Willett13a,Willett13b} but also some experiments that are not \cite{Lin12}.

Theoretical guidance can play an important role in identifying the state. Exact diagonalization
\cite{Morf98,Rezayi00,Peterson08,Peterson08b,Peterson12,Rezayi09,Wojs10,Storni10}
and density-matrix renormalization group \cite{Feiguin08,Feiguin09} studies of simplified model Hamiltonians show that non-Abelian states,
such as the Moore-Read (MR) Pfaffian state \cite{Moore91} and the anti-Pfaffian (aPf) state \cite{Lee07,Levin07} are viable ground states, but transitions to other
ground states can occur as a result of small changes in Hamiltonian parameters \cite{Rezayi00,Peterson08,Peterson08b,Wojs10}.
Since the details of the Hamiltonian matter (unlike in the case of states in the lowest Landau level, such as the $\nu=1/3$ state), it is important to
analyze Hamiltonians that model realistic experimentally-relevant systems and include
effects such as Landau level mixing and the finite-width of the quantum well.
Moreover, only a particle-hole symmetry-breaking effect, such as Landau level mixing, can split the degeneracy between
the MR Pfaffian and aPf states~\cite{Peterson08c,Wang09}.

The exact diagonalization study of 
Ref. \onlinecite{Wojs10} found the ground state for the half-filled $N=1$ Landau level for systems with up to $N_\Phi=2N_e - S = 33$ magnetic flux quanta 
in the spherical geometry using an effective Hamiltonian \cite{Bishara09a} that included Landau level mixing 
with virtual excitations to the $N=0$ and $N>1$ Landau
levels integrated out perturbatively to lowest-order in 
\begin{eqnarray}
\kappa = \bigl(\frac{e^2}{\epsilon\ell_0}\bigr)/{\hbar\omega_c}\propto 1/\sqrt{B}
\end{eqnarray}
 ($\ell_0=\sqrt{\hbar c/eB}$ is the magnetic length, $\epsilon$ is the dielectric constant of the host semiconductor, $\omega_c=eB/mc$ is the cyclotron frequency, and  
$S$ is a topological quantum number called the shift~\cite{Wen90b}).  The ground state at the MR Pfaffian shift of $S=3$ was found to have larger overlap with the MR Pfaffian wave function than the ground state
at the aPf shift of $S=-1$ had with the aPf wave function indicating, naively, that the ground state was in the MR Pfaffian universality class.
Two caveats are that: 1) Ref. \onlinecite{Wojs10} used two-body pseudopotentials \cite{Bishara09a} with a subtle normal-ordering error
that was corrected later \cite{Peterson13b,Sodemann13,Rezayi13} and 2) These results did not take into account the finite width of the quantum well.
Meanwhile, an exact diagonalization study \cite{Rezayi09} of a truncated Hamiltonian
for a few Landau levels found larger overlap with the aPf wave function on the torus.
(Similar ideas were used in Ref. \onlinecite{Wooten13}.) 
Ref.~\onlinecite{Rezayi09} used a truncated Hamiltonian approximation in hopes that it would capture the correct physics at intermediate values of $\kappa$,
even though it is uncontrolled, i.e., it is not exact in any limit,
unlike the Hamiltonians of Refs. \onlinecite{Bishara09a,Peterson13b,Sodemann13,Rezayi13}
which are exact in the $\kappa\rightarrow 0$ limit.
Moreover, the overlap between a ground state and a trial wave function may reflect
short-distance non-universal details of that particular trial wave function, rather than its universality class.
Indeed, such an overlap vanishes in the thermodynamic limit.

In this paper, we solve an effective Hamiltonian that incorporates both Landau level-mixing and finite quantum well width. We then
analyze the resulting ground states and low-lying excited states by several criteria. 
We begin by describing our effective Hamiltonian and providing a qualitative picture in Secs.~\ref{sec:model} and~\ref{sec:qualitative}.  In Sec.~\ref{sec:overlap} we 
compute the overlaps in the spherical geometry between the ground states at $S=3$ and $S=-1$
with, respectively, the MR Pfaffian and the aPf wave functions, and on the torus using the hexagonal unit cell where the MR Pfaffian and aPf occur at the same
flux and are orthogonal for an odd number of electrons.
We corroborate our overlap findings in Sec.~\ref{sec:ES} by calculating the entanglement spectrum.  In Sec.~\ref{sec:energy-gaps} we compare the
energy gaps in the spherical geometry at $S=3$ and $S=-1$
and provide estimates of the excitation gaps in the thermodynamic limit 
that take into account Landau level mixing and finite width.  In Sec.~\ref{sec:order-parameter} we introduce an operator that is odd under a particle-hole transformation and,
therefore, can be used as an order parameter distinguishing between the MR Pfaffian and aPf states.
We compute this order parameter in the ground state of our Hamiltonian on the torus and sphere.
According to all of these criteria, our central finding is that there is a ${\kappa_c}(w)$ such that 
the ground state for $0<\kappa < {\kappa_c}(w)$ is in the universality class of the MR Pfaffian.
We find that ${\kappa_c}(0)\approx 0.6$, monotonically increasing to ${\kappa_c}(4{\ell_0})\approx 1$.

A phase transition occurs at $\kappa = {\kappa_c}(w)$, identified by the collapse of both the energy gap
and the overlap with the MR Pfaffian wave function, as well as a sharp peak in the bipartite entanglement entropy.
For $w<1.5 \ell_0$, there appears to be a second phase transition at slightly larger $\kappa$. The intermediate
phase between the two transitions may be a different fractional quantum Hall state, such as the aPf
or a strong pairing phase \cite{Read00},
but the gap is too small for us to say anything definitive at these system sizes.
We culminate our findings in a phase diagram.

\section{Effective Hamiltonian}
\label{sec:model}

We diagonalize an effective Hamiltonian for spin-polarized electrons 
confined to the $N=1$ Landau level that incorporates the effects of Landau level mixing and finite width.  Finite width causes 
a `softening' of the Coulomb interaction at short distances and the Coulomb interaction can now cause virtual electron excitations 
to higher subbands of the quantum well in addition to higher Landau levels.  Hence, we take Landau level and subband mixing into account perturbatively to 
lowest-order in $\kappa$ through the terms that are generated by virtual excitations of 
 electrons to the $N=2,3, \ldots$ Landau levels and higher quantum well subbands or of holes to the $N=0$ Landau
level. As noted in Ref.~\onlinecite{Peterson13b}, virtual excitations into all unoccupied Landau levels are taken into account in this perturbative scheme producing a controlled model that is exact in the $\kappa\rightarrow 0$ limit.  This is in contrast to Landau level mixing models that work in an expanded, yet truncated, Hilbert space that are uncontrolled and not exact in any limit~\cite{Rezayi09,Papic2012,Zaletel14}.
Our effective Hamiltonian has the form:
\begin{eqnarray}
H(w/\ell_0,\kappa,N=1)&=&\sum_{m}V^{(2)}_{m}(w/{\ell_0},\kappa)\sum_{i<j}\hat{P}_{ij}(m)\nonumber\\
 &&\hspace{-1cm}+ \sum_{m} V^{(3)}_{m}(w/{\ell_0},\kappa)\sum_{i<j<k}\hat{P}_{ijk}(m)
\label{Heff}
\end{eqnarray}
where $\hat{P}_{ij}(m)$ and $\hat{P}_{ijk}(m)$ are projection operators that project, respectively,
the pair $i, j$ or triplet $i,j,k$ of electrons onto states of 
relative angular momentum $m$.  $V^{(2)}_{m}(w/{\ell_0},\kappa)$ 
and $V^{(3)}_{m}(w/{\ell_0},\kappa)$ are the 
two- and three-body effective pseudopotentials~\cite{Haldane83,Simon07c}, with
dependence on the well width $w/\ell_0$ and Landau level mixing parameter $\kappa$ denoted explicitly. 
In addition to the numerical renormalization of the two-body interaction, particle-hole symmetry breaking three-body terms are produced \footnote{More generally, at $k^{\rm th}$-order in $\kappa$, $k+2$-body interactions are generated.}. We only take into account the lowest $V^{(3)}_{m}$ for $m\leq 8$. Our results indicate that including higher three-body pseudopotentials has no effect on our conclusions.

In our calculations on the sphere, $N_e$ electrons are placed on a spherical surface of radius
$\sqrt{N_\Phi/2}$ with a radial magnetic field produced  by a magnetic monopole of strength
$N_\Phi/2$ at the centre ($N_\Phi/2$ is an integer or half-integer by Dirac's quantization condition).
Total angular momentum $L$ is a good quantum number, and any fractional quantum Hall state will be uniform and incompressible with 
 $L=0$~\cite{Haldane83,jain2007composite}.  For half-filling we have $N_\Phi=2N_e - S$; the filling factor in the $N=1$ Landau level 
 is given by $\nu=\lim_{N_e\rightarrow\infty} N_e/N_\Phi$.   As noted above, the MR Pfaffian has $S=3$ while the aPf has $S=-1$ which can be 
 seen by particle-hole transforming the MR Pfaffian.  

We model finite-width using both an infinite square well and Gaussian single-particle wave functions in the
$z$-direction, perpendicular to the two-dimensional electron gas.  We find that results for these two models are very similar and can be converted one into the other,
as we discuss further in Appendix \ref{sec:width-appendix}. Thus, we show only results for an infinite square
well.

We note that while we work in the spherical geometry, we utilize planar geometry pseudopotentials.  It has been argued that these more 
accurately represent the thermodynamic limit~\cite{Peterson08b}.  Furthermore we extrapolate several of our results to the thermodynamic 
limit and find that the choice of pseudopotentials is not crucial  (See Appendix~\ref{sec:finiteSize-sphericalpp}).

We also consider the torus geometry.  The torus is a two-dimensional 
plane with periodic boundary conditions with pseudomomentum $\mathbf{K}$ in a Brillouin zone that can be either rectangular or hexagonal.  On the torus 
there is no shift, and $N_\Phi = 2N_e$, which makes a direct comparison between MR Pfaffian and aPf states more
straightforward.

\section{Qualitative Picture}
\label{sec:qualitative}

There is very strong evidence that the ground state of Eq.~(\ref{Heff}) is in the MR/aPf universality class for $\kappa=0$ and that finite thickness increases 
the stability of this ground state~\cite{Morf98,Rezayi00,Peterson08,Peterson08b}.  This is true using the torus or spherical geometry.  A remaining question, 
and the one we answer here, is what happens under the influence of a particle-hole symmetry breaking effect like Landau level mixing, i.e., is the ground 
state in the MR or aPf universality class or neither universality class?

On the torus, at $\kappa=0$, the ground state is doubly degenerate in the thermodynamic limit
(over and above the $6$-fold topological degeneracy on the torus). One of these states is in the MR Pfaffian universality
class and the other is in the aPf universality class; their degeneracy is guaranteed by particle-hole symmetry.
On the sphere, the former occurs at $S=3$ and the latter at $S=-1$.
As $\kappa$ is increased, the two-body terms are modified and three-body terms are generated.
The former cannot break the symmetry between the MR and aPf states since they preserve particle-hole symmetry.

\begin{figure}[th!]
           \includegraphics[width=3.25in,angle=0]{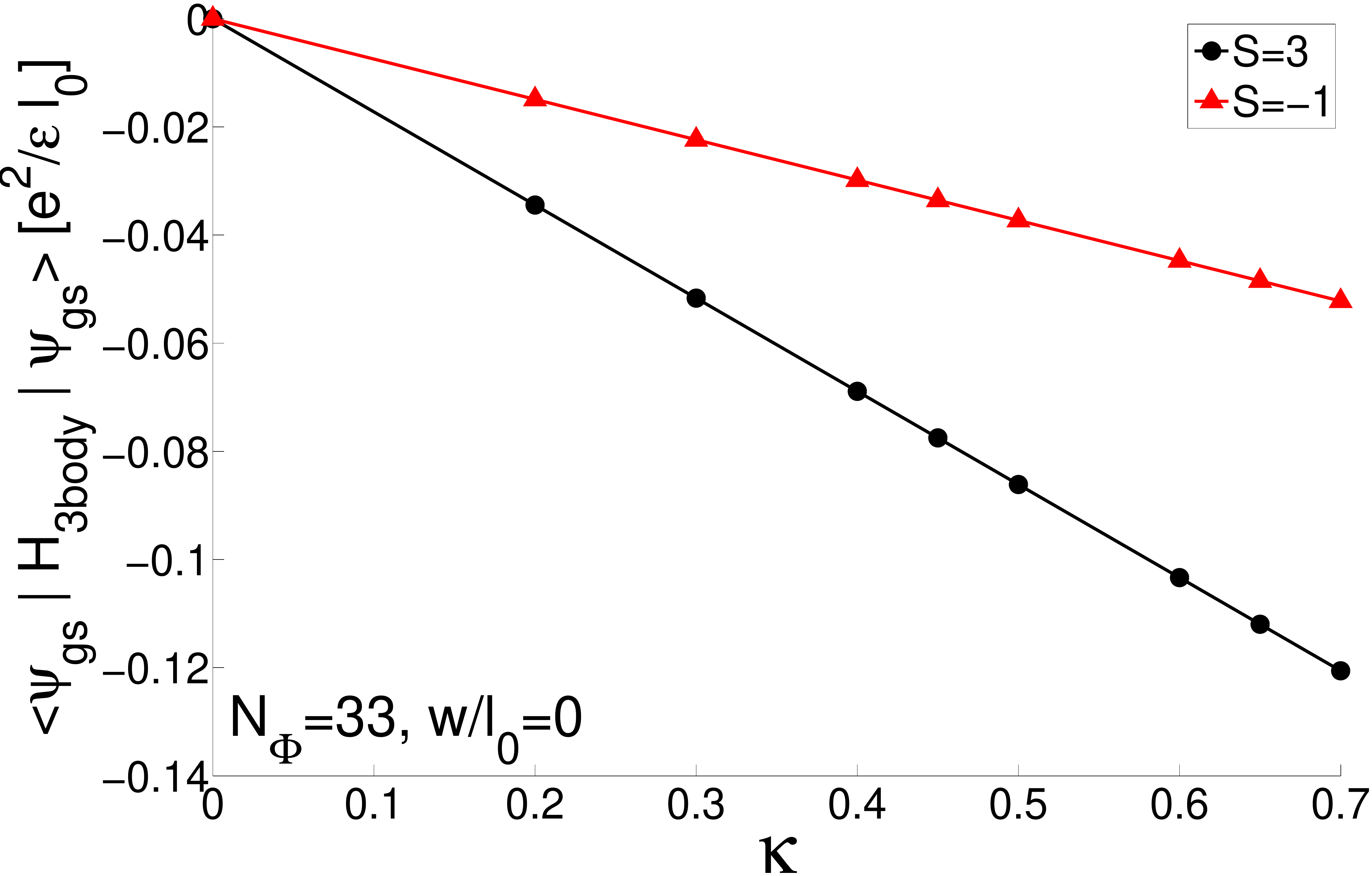}
    \caption{(Color online) Expectation value of the three-body terms in the Hamiltonian $\langle \Psi_\mathrm{gs}|H_\mathrm{3body}|\Psi_\mathrm{gs}\rangle$ in the $S=3$ (MR) and $S=-1$ (aPf) ground states obtained at $\kappa=0$ and $w/\ell_0=0$ in the system with $N_\Phi=33$ in the spherical geometry.  Here $H_\mathrm{3body}$ is the second term in Eq.~(\ref{Heff}) and introduces $\kappa$ dependence. This is the lowest-order perturbative contribution to the energies of these states. The energy of the $S=3$ (MR) state is lowered more.}
   \label{fig:H3-expectation}
\end{figure}

To understand the effect of the latter qualitatively, we consider their effect to lowest-order in perturbation theory,
i.e., we compute the expectation value of $H_\mathrm{3body}=\sum_m V_m^{(3)}(w/\ell_0,\kappa)\sum_{i<j<k}\hat{P}_{ijk}(m)$ in the two ground states on the sphere. As may be seen from Fig. \ref{fig:H3-expectation},
the energy of the $S=3$ state (MR) is lowered more than that of the $S=-1$ state (aPf).

\begin{figure}[h!]
\begin{center}
\includegraphics[width=8.25cm,angle=0]{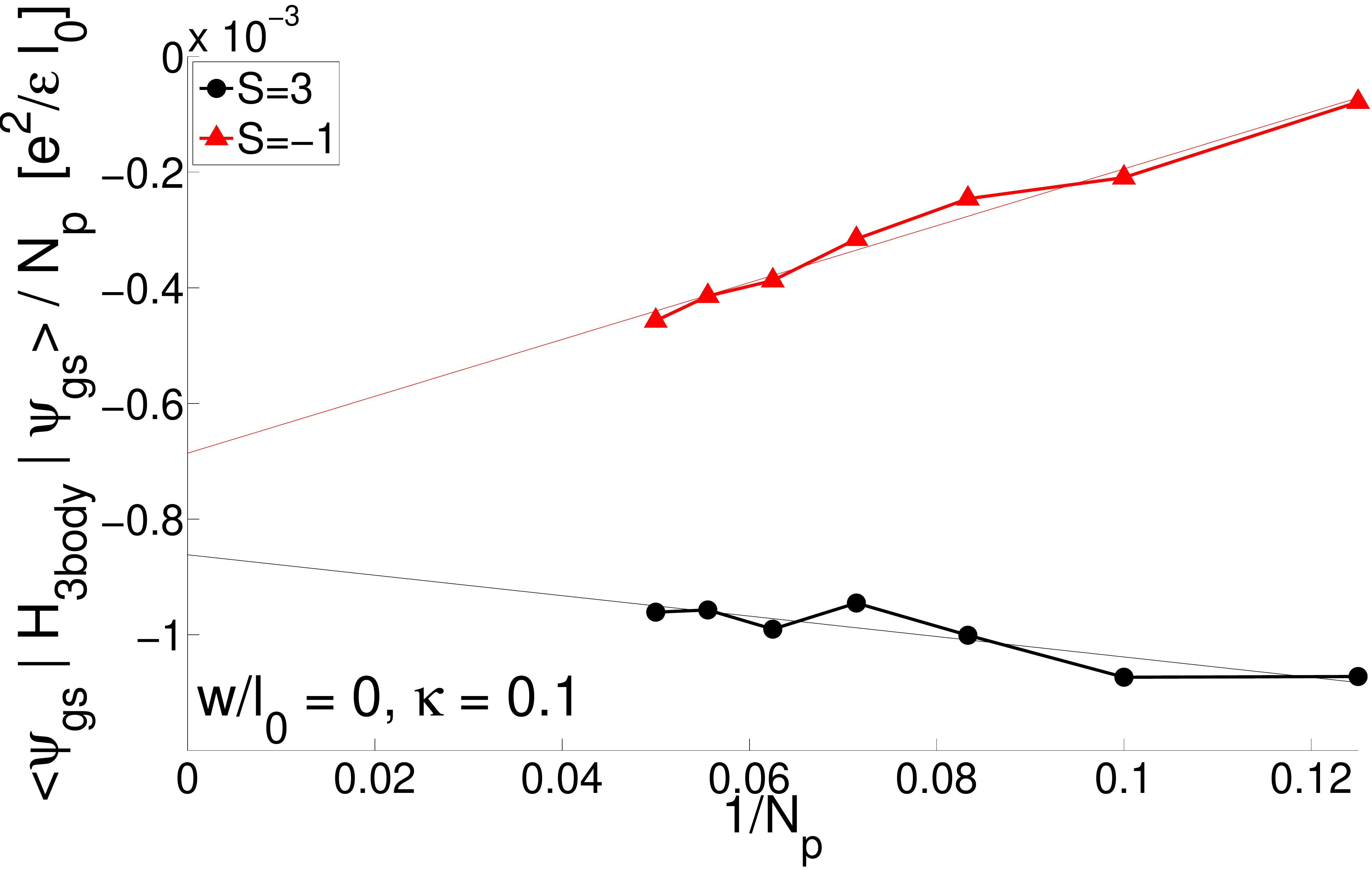}
\caption{The expectation value of the three-body terms in the Hamiltonian $\langle \Psi_\mathrm{gs}|H_\mathrm{3body}|\Psi_\mathrm{gs}\rangle$ per particle at various system sizes. This is the leading contribution to the energy difference between these states computed
perturbatively in $H_\mathrm{3body}$. $N_p$ is the number of electrons for $S=3$ and number of holes for $S=-1$ .
 The energy difference between the extrapolated values is 0.00018 $e^2/\epsilon\ell_0$ = 0.12 $\kappa |V^{(3)}_3|$.
 }
\label{fig:afs_avgH3_extrapolated}
\end{center}
\end{figure}

The preceding calculation was done at $N_\Phi = 33$.
To check whether this conclusion is likely to hold in the thermodynamic limit,
we repeat it for different system sizes and consider the extrapolation to $N_e=\infty$.
In Fig.~\ref{fig:afs_avgH3_extrapolated}, we plot the expectation value per particle of the three-body terms of $H(w/\ell_0=0, \kappa=0.1, 1)$ given in Eq.~(\ref{Heff}) evaluated in the Coulomb ground state for systems with $N_{\Phi}=13$ to $N_{\Phi}=37$. A linear fit in the inverse number of particles/holes provides an estimate for the thermodynamic limit. We observe that the energy at $S=3$ is lowered more than at $S=-1$ for all available system sizes as well as in the thermodynamic limit.  This is in agreement with our conclusions drawn in previous paragraph and in
Fig. \ref{fig:H3-expectation}.
Thus, from this result, we expect the MR state to be the ground state
for small $\kappa$. We verify this expectation by exact diagonalization in the sections that follow.

\begin{figure}[ht!]
        \includegraphics[width=7.7cm,angle=0]{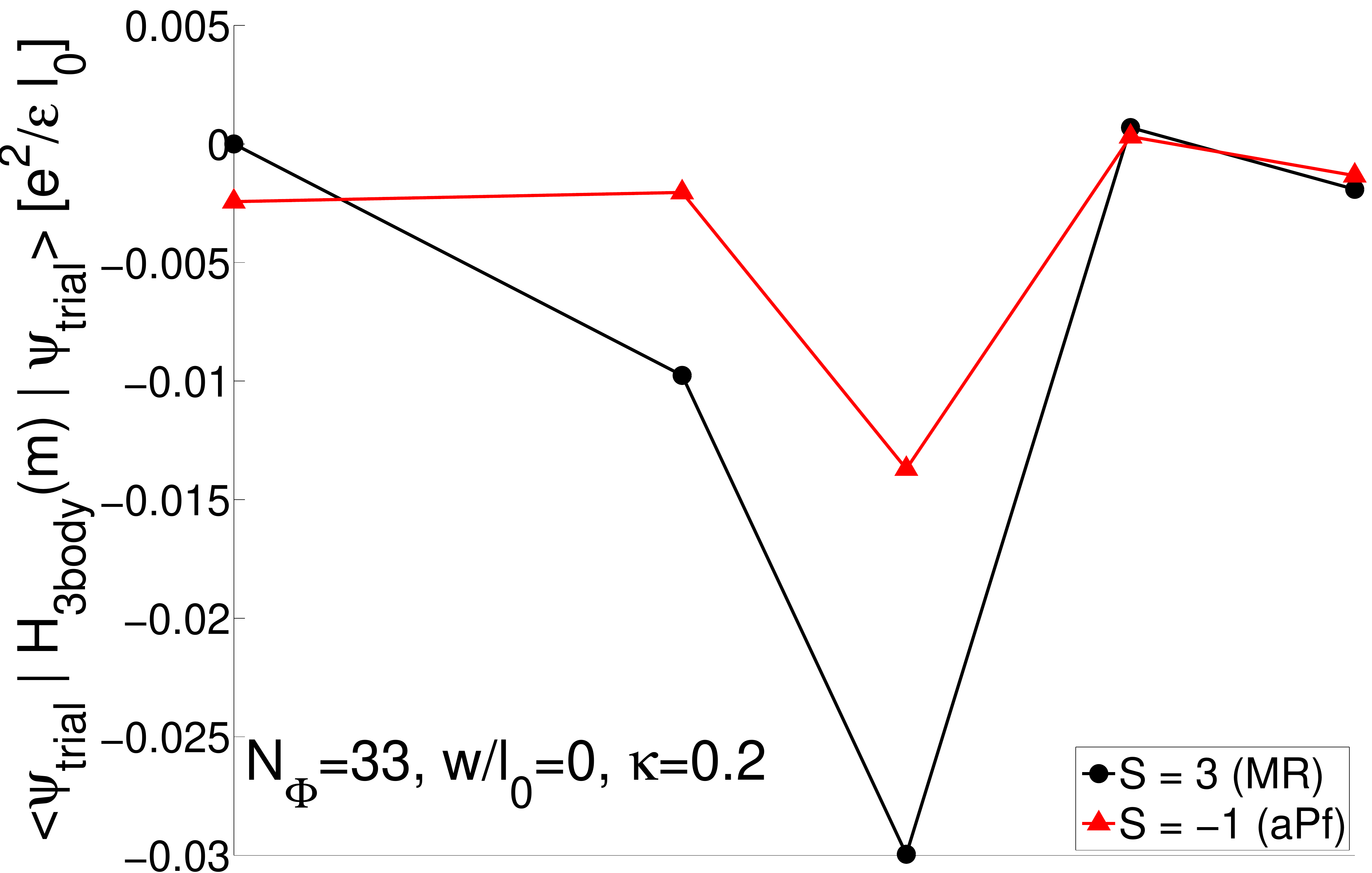}\\
\includegraphics[width=7.7cm,angle=0]{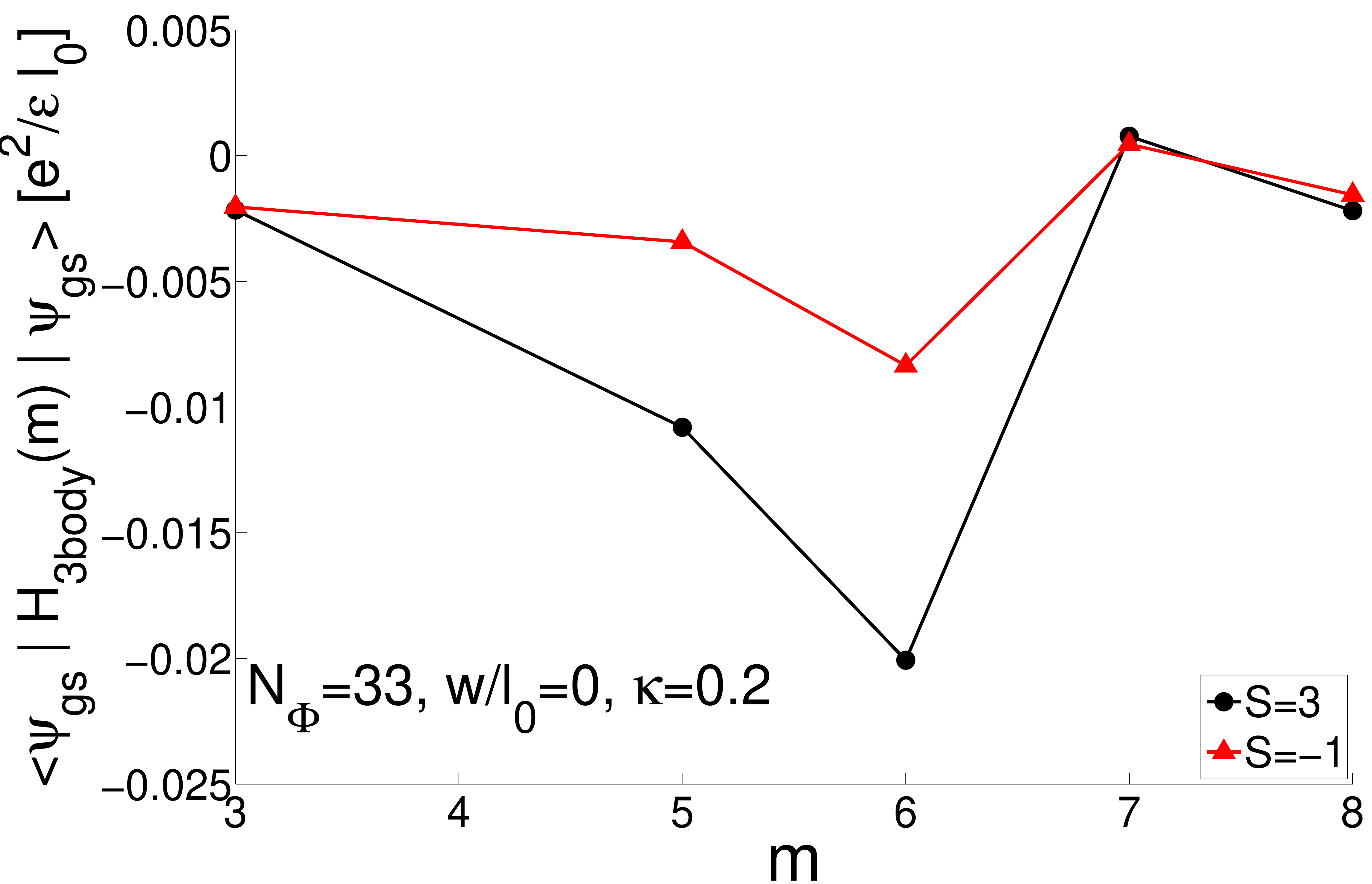}
    \caption{(Color online) Top Panel: Expectation value of the three-body terms in the MR and aPf trial
wavefunctions as a function of angular momentum $m$, i.e., $\langle \Psi_\mathrm{trial}|H_\mathrm{3body}(m)|\Psi_\mathrm{trial}\rangle$ where $H_\mathrm{3body}(m)=V^{(3)}_m(w/\ell_0,\kappa)\sum_{i<j<k}\hat{P}_{ijk}(m)$, and $\Psi_\mathrm{trial}=\Psi_\mathrm{MR}$ or $\Psi_\mathrm{aPf}$.
Bottom Panel: Expectation value of the three-body terms for the  $S=3$ and $S=-1$ ideal Coulomb ground states
as a function of angular momentum $m$. }  
   \label{fig:H3-by-L}
\end{figure}

The manner in which the three-body terms favor the MR state is subtle.
The lowest angular momentum
term, $V^{(3)}_{3}(w/{\ell_0},\kappa)$, has vanishing expectation value
in the MR trial wave function and small but negative expectation
value in the aPf trial wave function and, therefore, one might expect the aPf state to have lower energy if
$V^{(3)}_{3}(w/{\ell_0},\kappa)$ dominates over higher angular momenta.  However, 
as may be seen from
the top panel of Fig. \ref{fig:H3-by-L},
the energy contributions of the $V^{(3)}_{m}(w/{\ell_0},\kappa)$ for $m=5$ and 6 are
generally larger and will dominate (we have chosen $w/\ell_0=0$ and $\kappa=0.2$ for illustrative purposes).   Of course, the MR wave function has 
a vanishing expectation value of $V^{(3)}_{3}(w/{\ell_0},\kappa)$ since this operator completely annihilates the wave function, i.e., the MR wave function 
is the zero-energy ground state of $\sum_{i<j<k}\hat{P}_{ink}(3)$ from which $V^{(3)}_{3}(w/{\ell_0},\kappa)$ is constructed.  The aPf wave function has 
a nearly zero expectation value of $V^{(3)}_{3}(w/{\ell_0},\kappa)$ because at $S=-1$ we can particle-hole transform $V^{(3)}_{3}(w/{\ell_0},\kappa)$ to 
give a three-body operator that exactly annihilates the aPf wave function but also produces two-, one-, and zero-body terms.  Thus,  $V^{(3)}_{m}(w/{\ell_0},\kappa)$ for $m>3$ terms will largely determine which state has lower energy.  Moreover, the above expectation remains when we use the actual
Coulomb ground state, again for $w/\ell_0$ and $\kappa=0.2$, 
rather than the trial wave functions.  Then 
we find that the energy difference due to $V^{(3)}_{3}(w/{\ell_0},\kappa)$ becomes negligible and the relative importance of the higher angular momenta is enhanced, as may be seen in the bottom panel of Fig. \ref{fig:H3-by-L}.  Hence, the effect of the three-body terms due to Landau level mixing and finite width cannot be simply modelled by considering only the lowest three-body relative angular momentum ($m=3$) term -- this is similar to how the effect of finite width alone cannot be completely understood by simply looking at the ratio of the $m=1$ and $m=3$ Haldane pseudopotentials~\cite{Peterson08b}.

\section{Wave Function Overlap}
\label{sec:overlap}

According to the argument of the previous section, the ground state is in the MR universality class for small $\kappa$.
We now corroborate the arguments of the previous section using wave function overlap.

The MR  wave function takes the following form on the sphere:
\begin{equation}
\label{eq:Pf-trial-wavefunction}
\Psi_\text{MR} = \text{Pf}\left(\frac{1}{{u_i}{v_j} - {v_i}{u_j}}\right) \prod_{i>j} \left({u_i}{v_j} - {v_i}{u_j}\right)^2
\end{equation}
where $({u_i}, {v_i}) = (e^{-i{\phi_i}/2}\cos{\theta_i}, e^{i{\phi_i}/2}\sin{\theta_i})$
are the spherical coordinates of the $i^{\rm th}$ particle. Here Pf denotes the Pfaffian, i.e., the square root of the determinant of
an antisymmetric matrix. On the torus, this wave function takes the form:
\begin{equation}
\Psi_\text{MR} = \text{Pf}\left(\frac{{\vartheta_a}({z_i}-{z_j})}{{\vartheta_1}({z_i}-{z_j})}\right)
\prod_{i>j} \left({\vartheta_1}({z_i}-{z_j})\right)^2
\, \Phi_{\rm c.m.}\!\big(\mbox{$\sum_i {z_i}$}\big)
\end{equation}
Here, ${\vartheta_1}(z)$ and ${\vartheta_a}(z)$, $a=2,3,4$ are the Jacobi theta functions
and $\Phi_{\rm c.m.}\!\big(\mbox{$\sum_i {z_i}$}\big)$ is the center-of-mass wave function.  $z_{i}$ is a complex planar coordinate of the $i^{\rm th}$ particle.
The aPf wave functions on the sphere and torus are obtained by taking the particle-hole conjugates of these wave functions.

\begin{figure}[tbh!]
            \includegraphics[width=4.27cm,angle=0]{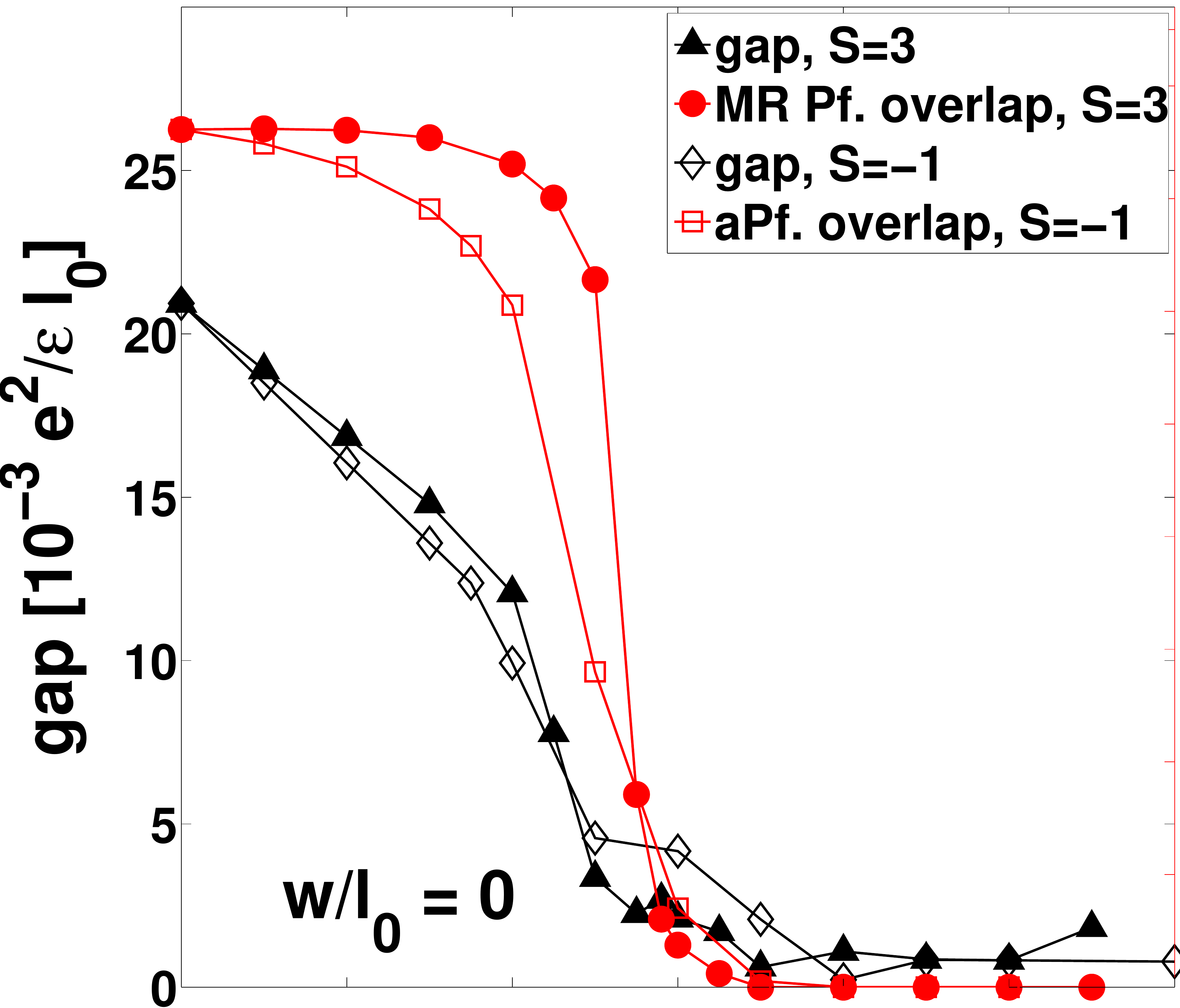}
           \includegraphics[width=4.24cm,angle=0]{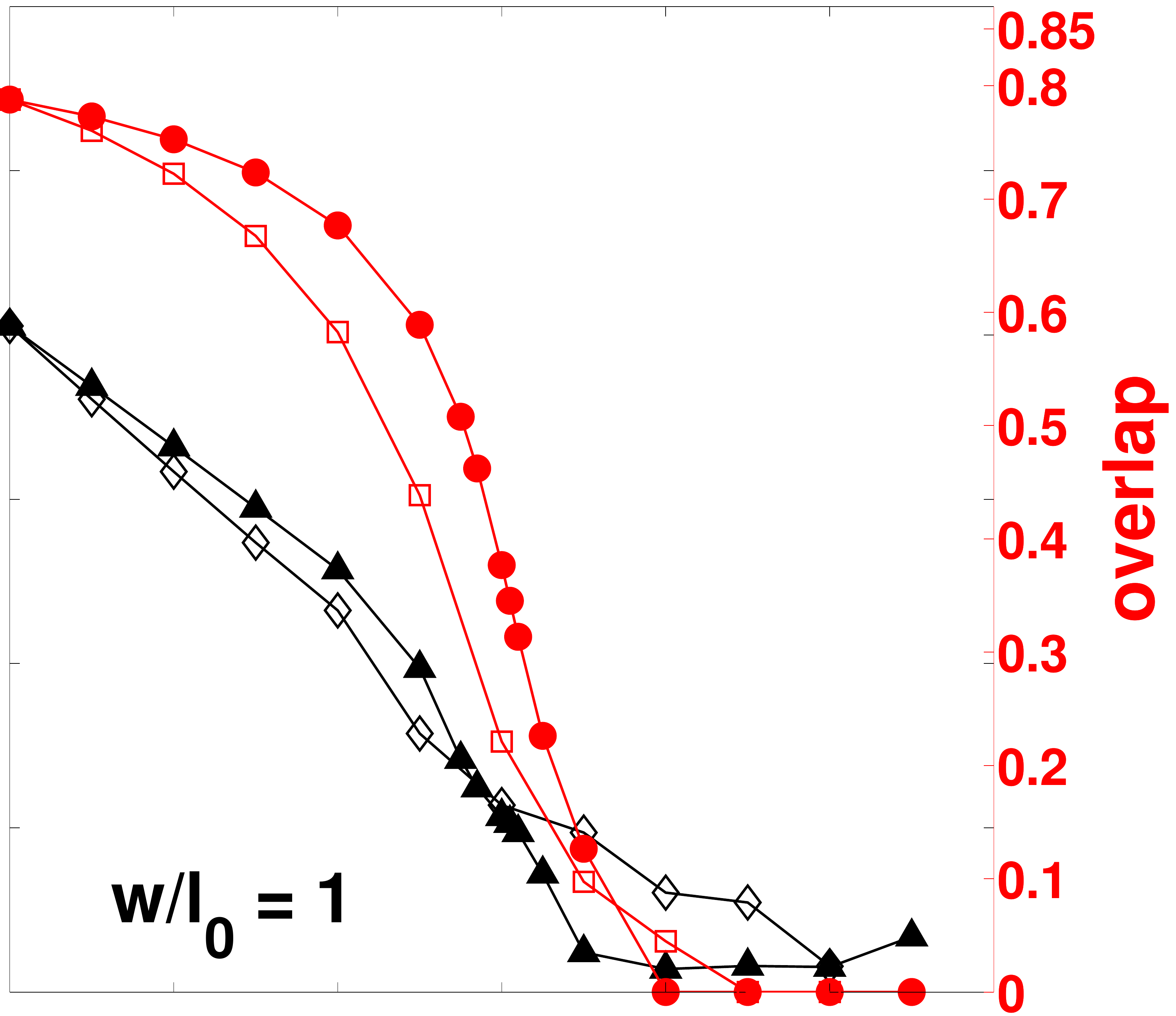}\\
            \includegraphics[width=4.27cm,angle=0]{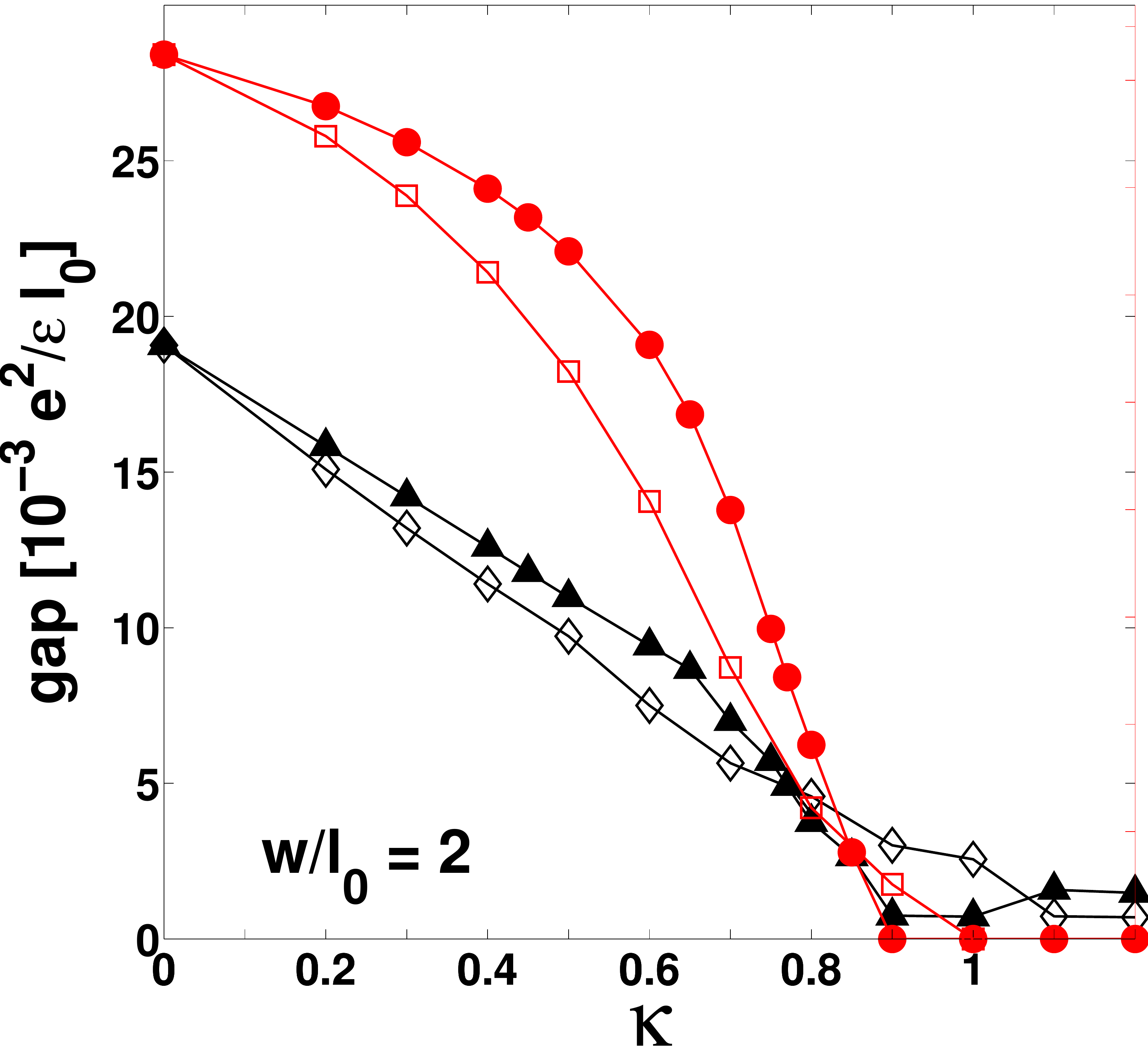}
            \includegraphics[width=4.15cm,angle=0]{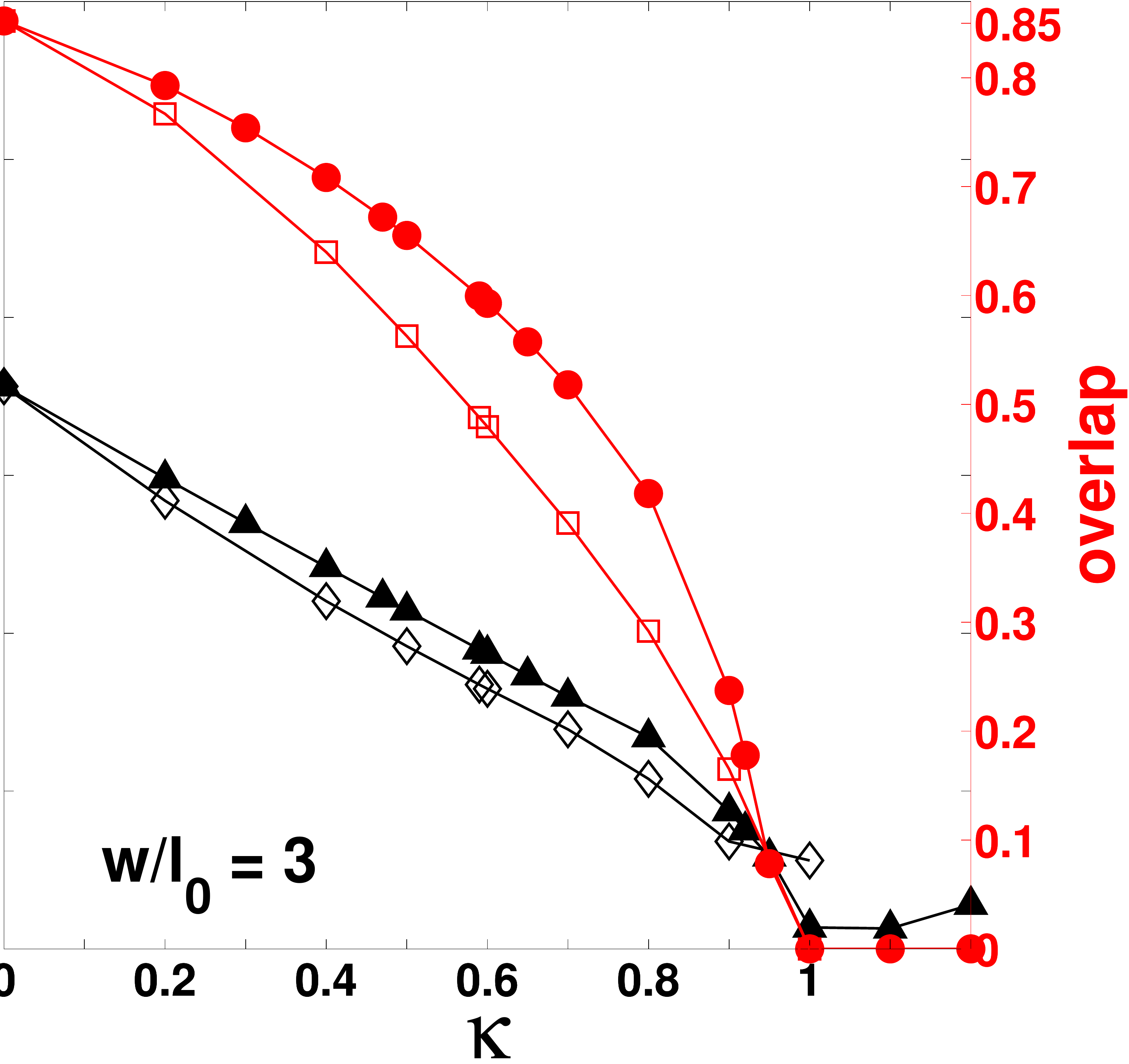}
      \caption{
        (Color online) Energy gap (energy difference between the two lowest states)
and model wave function overlap ($N_\Phi=33$ system) at the MR Pfaffian (${N_e}=18$)
and aPf (${N_e}=16$) shifts.  Note that for small $\kappa$ both the gap and overlaps are higher for the MR Pfaffian than they are for the aPf.
     }
  \label{fig:overlap-and-gap}
\end{figure}

Fig.~\ref{fig:overlap-and-gap} shows the numerical wave function overlaps between the ground state at $S=3$ and $S=-1$ for ${N_\Phi}=33$ and, respectively,
the MR  and aPf wave functions on the sphere as a function of $\kappa$
for $w/{\ell_0}=0,1,2$, and 3.   An overlap of unity or zero means the exact ground state of Eq.~(\ref{Heff}) is either identical 
to or completely different from the trial MR or aPf wave function.  We remind  that an overlap is not a universal quantity of a ground state 
that can be extrapolated to the thermodynamic limit since, unless it is unity for all $N_e$, it will vanish as the number of particles goes to infinity.
The overlap between the ground state of Eq.~(\ref{Heff}) and both the MR and aPf wave functions are reasonably large for small $\kappa$ and drop dramatically at larger $\kappa$,
falling to zero somewhere in the range $0.7 - 1.0$, with larger $\kappa$ occurring for larger widths.
Importantly, the overlap with the MR state is consistently larger.  Although not shown here, smaller system results are consistent with the $N_\Phi=33$ results.  
This is an indication that the ground state is likely to be in the same universality class as the MR
state for small $\kappa$.  But, as we cautioned above, it is possible that the aPf's smaller overlaps are merely expressing
the fact that non-universal short-distance physics is not well-captured by this wave function.

On the torus, the MR and aPf states occur at precisely the same flux.  With a rectangular unit cell the MR and aPf states are 3-fold degenerate (after factoring out the 2-fold center-of-mass degeneracy) with each zero-energy state existing at $\mathbf{K}=(0,N_0/2), (N_0/2,0),$ and $(N_0/2,N_0/2)$ where $N_0$  
is the greatest common divisor of $N_e$ and $N_\Phi$.  $K_x$ and $K_y$ are in units of $2\pi \hbar/a$ and $2\pi \hbar/b$ where $b/a$ is the aspect ratio of the rectangular unit cell.  Generically, in this geometry, the MR and aPf are not orthogonal rendering ambiguous the use of overlaps.
However,
in the hexagonal unit cell containing an odd number of electrons, the MR and aPf states are orthogonal and both have $\mathbf{K}=(0,0)$. 
At $\kappa=0$ the Coulomb ground state is a
doublet at $\mathbf{K}=(0,0)$ (provided $N_e\neq 6n+1$)
and we find that for nonzero $\kappa$ this doublet is split in such a way that each member has a nonzero
overlap with either MR or aPf state, as described by Papic \textit{et al}.~\cite{Papic2012}. 
The lowest-lying state has nonzero overlap only with the MR state.

\begin{figure}[th!]
           \includegraphics[width=8.5cm,angle=0]{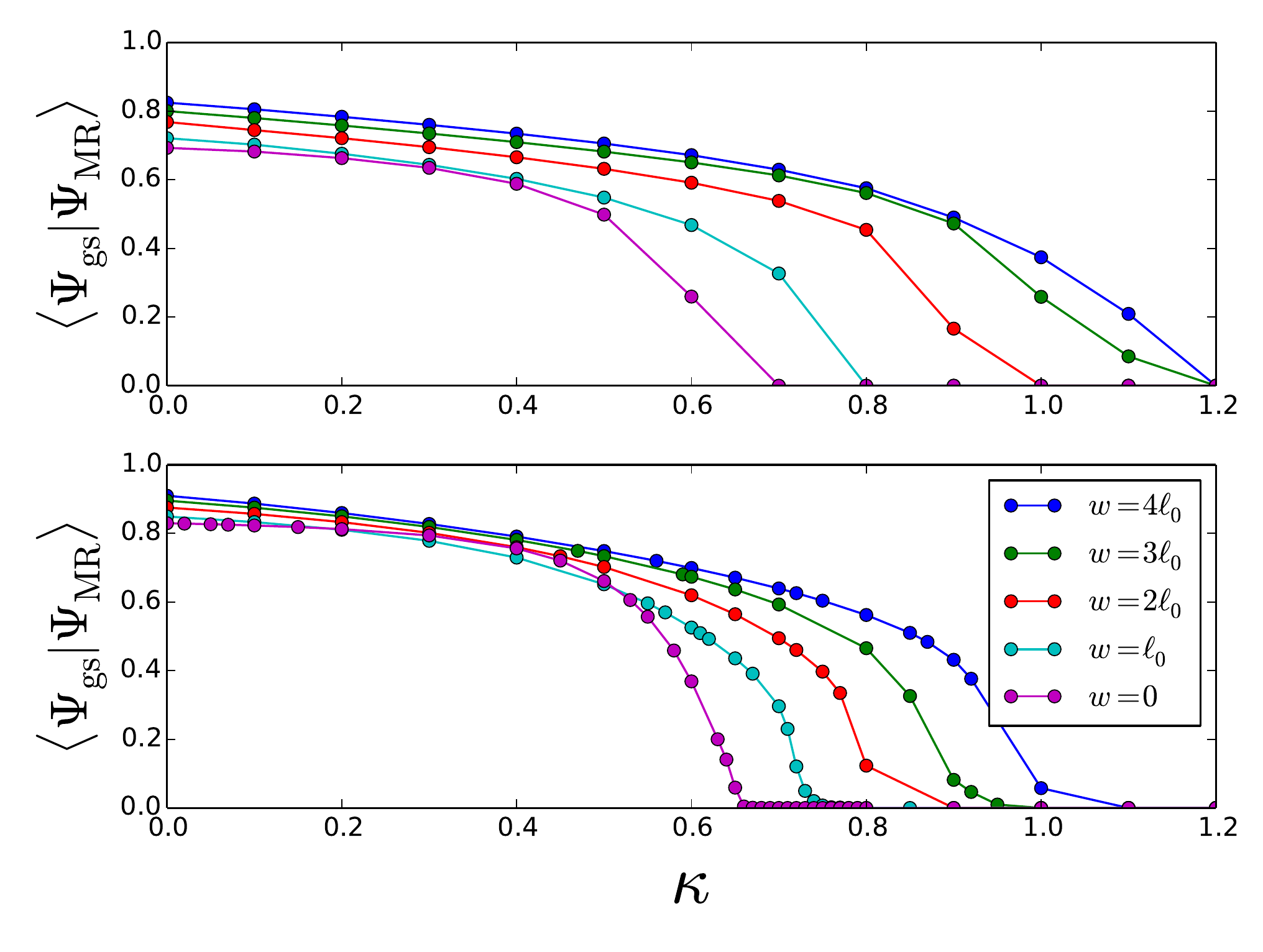}
    \caption{(Color online)
The overlap between the ground state and the MR wave function as a function
of $\kappa$, for $w/\ell_0=0,1,2,3$, and 4.  $w/\ell_0=0$ is the left-most curve and $w/\ell_0=4$ is the right-most. Top panel is
 on the torus for ${N_e}=15$ and hexagonal unit cell, bottom panel for the sphere with $N_\phi=29$ with $S=3$ ($N_e=16$).
     }
  \label{fig:torus-overlaps}
\end{figure}

The top panel of Fig.~\ref{fig:torus-overlaps} shows the overlap between the MR state and the 
ground state for the hexagonal unit cell as a function of $\kappa$ and $w/\ell_0$ for $N_e=15$.  The overlap is relatively large, dropping to zero at a critical
$\kappa$ in the range $0.6 -1$, with larger values occurring for larger widths.  
Meanwhile, on the torus the first excited state has a similarly large overlap with the aPf wave function, essentially mirroring the overlap between the ground state and the MR wave function. 
The overall shape of the overlap is very similar to that on the sphere, shown in the bottom panel of Fig.~\ref{fig:torus-overlaps},
further corroborating previous results and, as we show below, these conclusions are supported by criteria that
do not depend on any particular trial wave functions.

Our results for $S=-1$ on the sphere and for the first excited state on the torus are a bit surprising.
If the ground state at $S=3$ is firmly in the MR universality class, then the ground state at $S=-1$ should have 8 quasiholes
on the $S=3$ ground state. Instead, it has high overlap with the aPf state. Similarly, the first excited state on the torus should
look like an exciton on the MR ground state but, instead, it has high overlap with the aPf ground state. If the ground state is in the universality class of the MR state, then the energy gap to a state with high overlap with the aPf
should be extensive in system size. What we observe can thus only happen in small systems. For larger system sizes, the $S=-1$ ground state
on the sphere and the first excited state on the torus must look, respectively, like the $S=3$ ground state on the
sphere or the torus with excitations on top.

\section{Entanglement Spectrum}
\label{sec:ES}

We have called the $S=3$ and $S=-1$ 
ground states the MR state and the aPf state, respectively, due to their large
overlaps with the corresponding trial wave functions (Eq.~(\ref{eq:Pf-trial-wavefunction}), 
and its particle-hole conjugate). However, the overlap with trial wave functions
is not universal and vanishes in the thermodynamic limit. Therefore, we now
identify these states by a universal criterion, the entanglement spectrum.

In the spherical geometry, we divide the system in two pieces A and B~\cite{Levin06,Kitaev06b,Haque07,Zozulya07,Biddle11}, and 
obtain the reduced density matrix for one half by tracing out the degrees of freedom of the remaining half.  
The eigenvalues $\rho_n$ of the density matrix are interpreted as energies, $\rho_n \equiv e^{-\xi_n/2}$ \cite{Li08}.
If we make a cut in orbital space, then the entanglement spectrum for a state in the MR universality class should
have negative slope for the entanglement energies as a function of the $z$-component of the angular momentum $L_z^A$ in 
sector $A$, for example, as discussed in Ref.~\onlinecite{Li08}.  

\begin{figure}[th!]
	\includegraphics[width=7.0cm,angle=0]{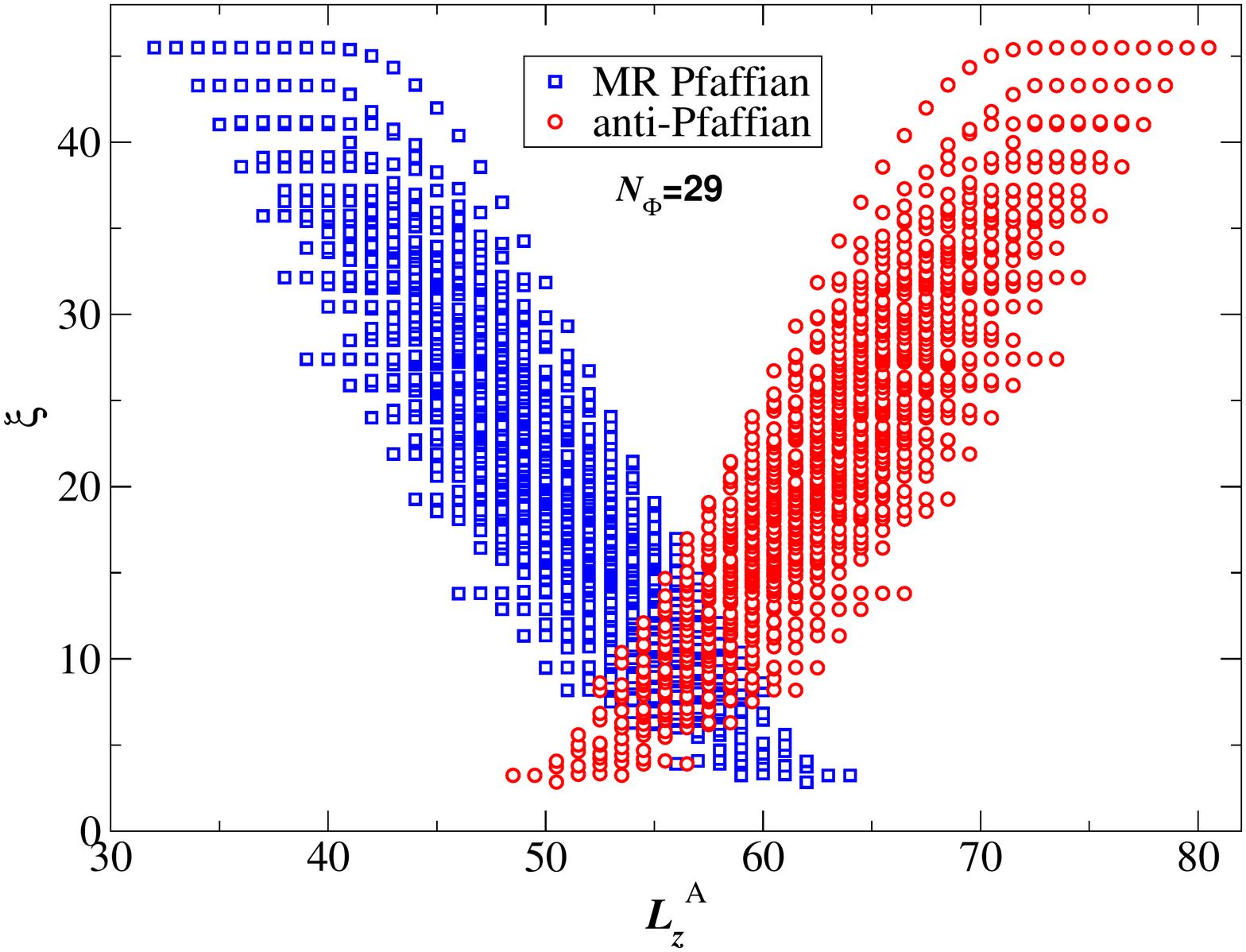}
	\includegraphics[width=7.0cm,angle=0]{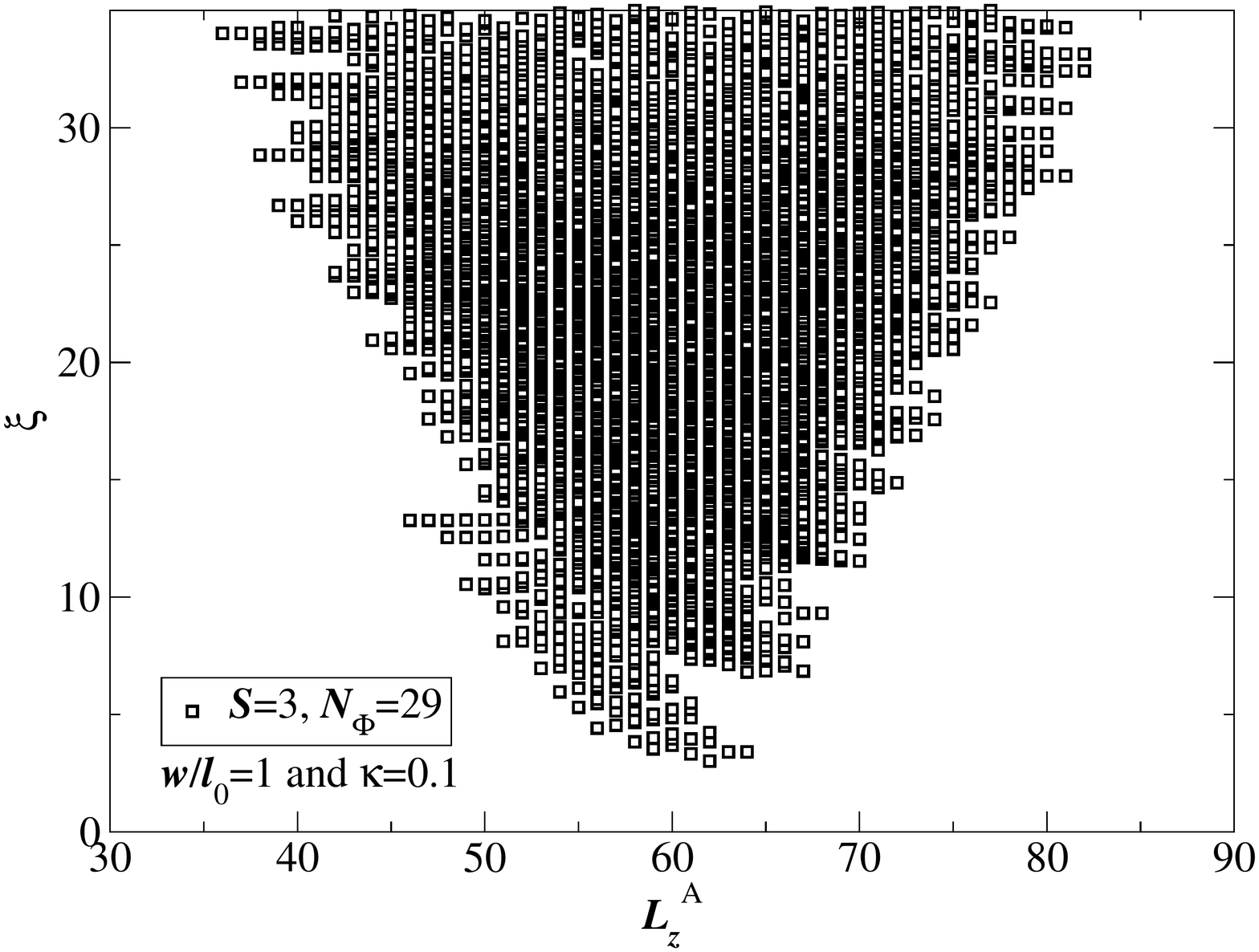}
	\includegraphics[width=7.0cm,angle=0]{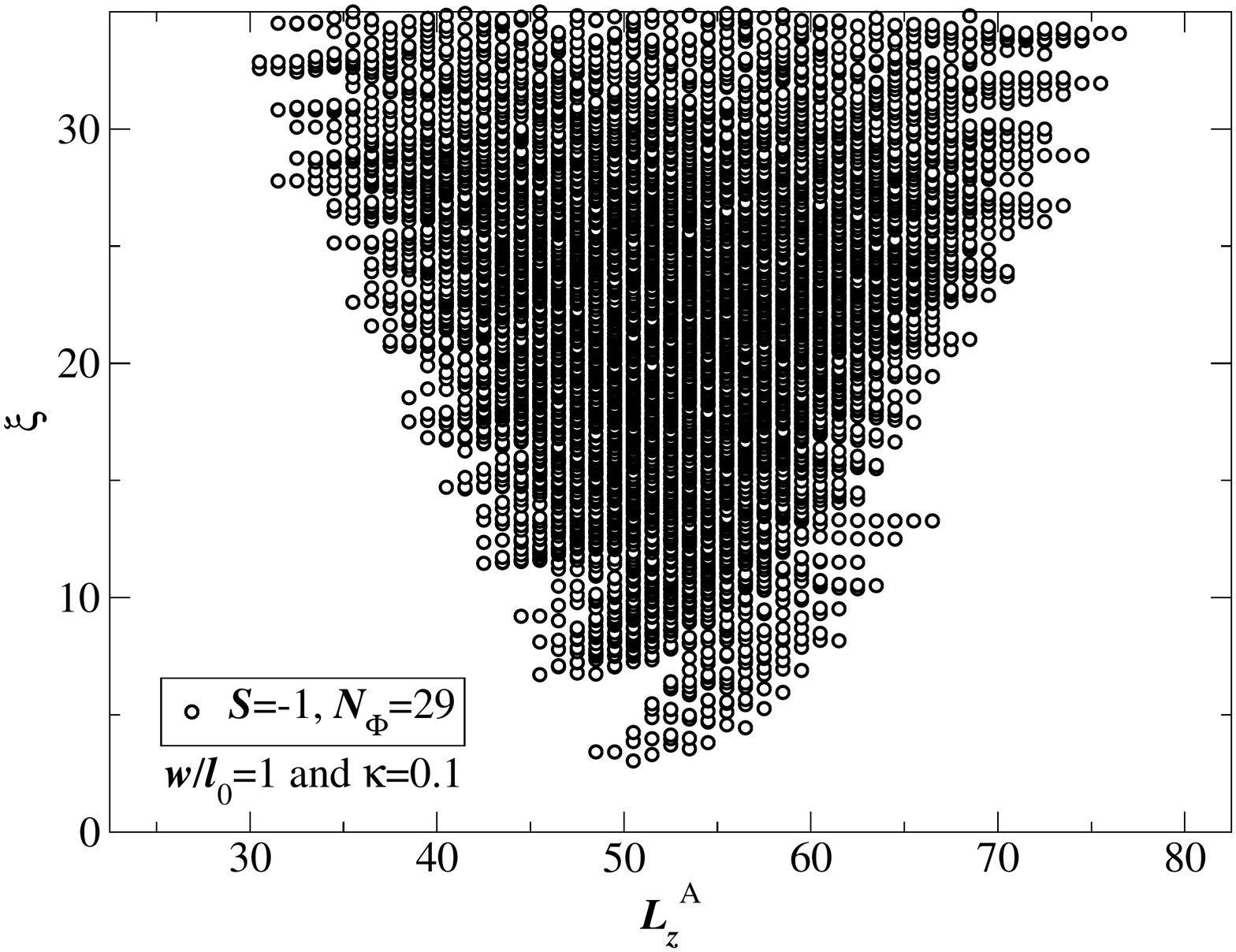}
          \caption{(Color online) The entanglement spectrum for the MR Pfaffian (blue) and aPf (red) on the sphere at $S=3$ and $S=-1$, respectively, 
    are shown in the top panel.  The MR state has a negative slope (discussed more in the text) while 
    the aPf has a positive slope.  The middle and lower panel show the entanglement spectrum for the exact ground state at $S=3$ 
    and $S=-1$, respectively, for  $\kappa=0.1$ and $w/\ell_0=1$.  At $S=3$, the low-lying states show the same slope and level structure
   as the MR state and at $S=-1$ they show the same slope and level structure as the aPf state.  Both systems are at $N_{\Phi}=29$ 
   and were partitioned to have 15 orbitals in each hemisphere (this corresponds to the partition of $P[0|0]$ in 
Li and Haldane's notation~\cite{Li08}).}
  \label{fig:ES}
\end{figure}

A state in the MR Pfaffian universality class displays the following structure in the entanglement spectra:  the spectra is essentially divided 
into two pieces by a ``gap" with the low-lying states corresponding to the conformal field theory (CFT) 
describing the MR edge states.  Starting from the ``root" configuration
 of the MR states one can define  $\Delta L_z^A = (L_z^A)_\mathrm{root}-L_z^A$ where  $(L_z^A)_\mathrm{root}$ is the $z$-component of 
 angular momentum of the ``root" configuration, cf. Ref.~\onlinecite{Li08}.  
The slope of the ``energy" spectra, i.e., whether  $\Delta L_z^A$ is positive or negative as a function of $L_z^A$, expresses 
the chirality of the edge modes 
of the CFT.  In our convention, a state in the MR universality class has an entanglement spectra  with a negative slope.  Thus, the 
entanglement spectrum for a state in the aPf universality class has a positive slope corresponding to edge modes with opposite chirality.

\begin{figure}[bht!]
\begin{center}
  \includegraphics[width=8.5cm,angle=0]{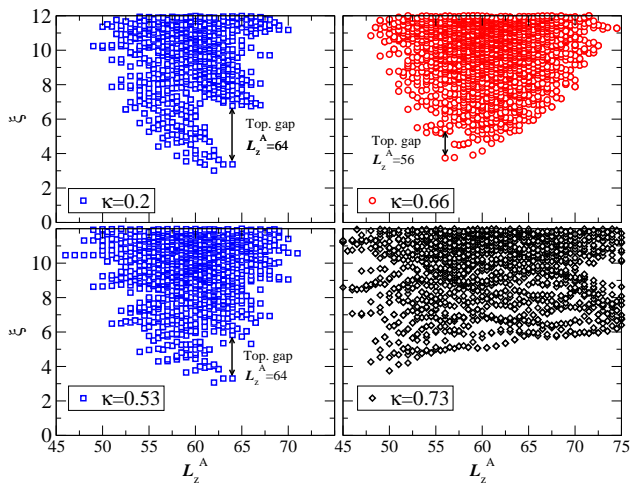}
  \caption{(Color online) Entanglement spectrum for $N_\Phi=29$ at $\kappa=0.2$, 0.53, 0.66, and 0.73.  
The entanglement spectra at $\kappa=0.2$ and 0.53 are shown in blue to emphasize their consistency with
 a ground state in the universality class of the MR Pfaffian. At $\kappa=0.66$ we show the entanglement spectra between the
  two entanglement entropy peaks in the lower panel of Fig.~\ref{EE-peaks}. It is colored red to indicate that its low-lying
  level structure has some similarity to states in the aPf universality class although the entanglement gap is too
small to allow for any definitive statements.
At $\kappa=0.73$ the entanglement spectra completely changes to that of the unknown phase to the right of the second entanglement entropy peak in Fig.~\ref{EE-peaks}.  We also indicate the definition of the ``topological gap'' for $\kappa=0.2$, 0.53, and 0.66 as described in the text.}
\label{fig-ES-all}
\end{center}
\end{figure}

Fig. \ref{fig:ES} shows that the entanglement spectrum at $S=3$ for $\kappa=0.1$ and $w/\ell_0=1$
has negative slope, similar to that of the entanglement spectrum for the MR trial wave function, Eq.~(\ref{eq:Pf-trial-wavefunction}).
Meanwhile the entanglement spectrum at $S=-1$ has positive slope, similar to that of the aPf trial
wave function (the particle-hole conjugate of Eq.~(\ref{eq:Pf-trial-wavefunction})). 
We therefore find that both the entanglement spectrum and  
overlaps allow us to identify the $S=3$ and $S=-1$ ground states as the MR state and the aPf state, respectively.
The phase transition at $\kappa\approx 0.6-1.0$ is also observed in the entanglement spectra,
as shown in Fig. \ref{fig-ES-all} and discussed further in Sec.~\ref{sec:phase-diagram}.  As $\kappa$ increases, the structure of the low-lying states first changes chirality and then changes completely and no longer resembles the MR or aPf entanglement spectra.

We adopt the definition of the ``topological gap" for the Pfaffian-like phase introduced in~\cite{Li08} with $\Delta L_z^A=0$, thus defining it as difference between the single universal level at $L_z^A=64$ and the lowest generic level at the same $L_z^A$ (see Fig.~\ref{fig-ES-all}).  In addition we track the difference between the lowest two levels at $L_z^A=56$, which is the symmetry point between the MR Pfaffian and aPf spectra (see the Top Panel of Fig.~\ref{fig:ES}) and also appears to be the lowest-$L_z^A$ ``universal" level after the first phase transition ($\kappa\approx0.66$). 

\begin{figure}[bht!]
\begin{center}
  \includegraphics[width=8.5cm,angle=0]{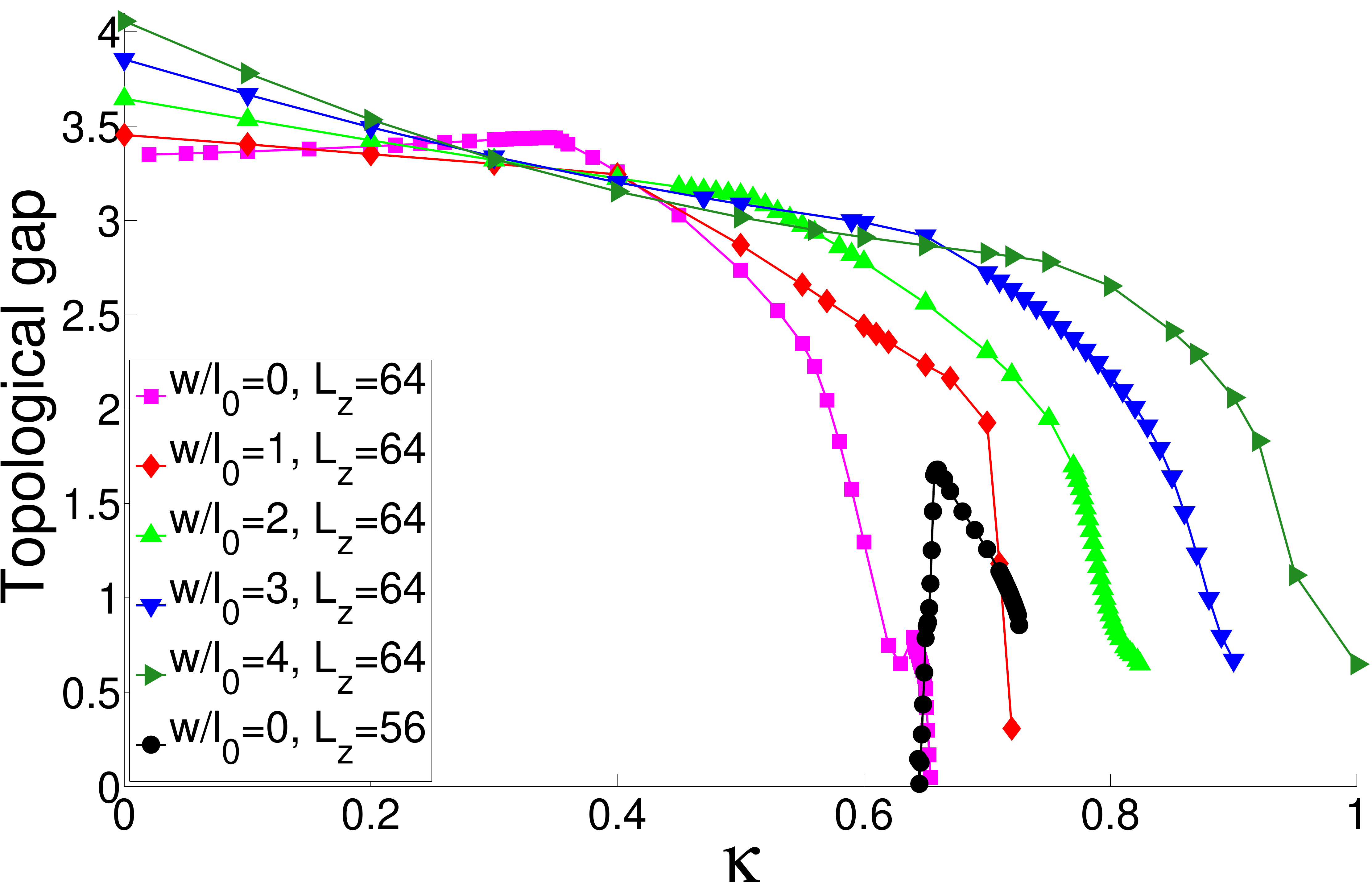}
  \caption{(Color online) Topological gap, the difference between the universal and lowest generic level at $L_z^A=64$, is calculated in spherical geometry at $S=3$ for the system with $N_{\Phi}=29$.  Note that qualitatively the behaviour and collapse of the topological gap is similar to the collapse of the overlaps shown in Fig.~\ref{fig:torus-overlaps}.  We also show the
  topological gap at $L_z^A=56$ (black filled circles) in the region of parameter space where the entanglement spectrum qualitatively changes and has the same chirality as the aPf state (cf. the $\kappa=0.66$ panel in Fig.~\ref{fig-ES-all}). }
\label{fig-top_gap_S3}
\end{center}
\end{figure}

In Fig.~\ref{fig-top_gap_S3} we show the topological gap in the Pfaffian-like phase for different widths along with the $L_z^A=56$ gap at $w/\ell_0=0$. We observe that the topological gap remains relatively robust to the variations of finite thickness and Landau level mixing strength for small $\kappa$. For each width there exists a critical value of $\kappa$, that can be approximately inferred from the MR Pfaffian overlap, where the topological gap vanishes. We see that the $L_z^A=56$ gap displays a sharp jump simultaneous with the vanishing of the MR Pfaffian topological gap at $w=0$. This may indicate a topological phase transition where the new state is also topological but has opposite chirality. We further discuss this state in the Section~\ref{sec:phase-diagram}. With increasing Landau level mixing strength the $L_z^A=56$ gap is suppressed until a different phase appears around $\kappa=0.73$.   

We also studied the dependence of the MR Pfaffian topological gap on the system size for $N_{\Phi}$ up to 29 (not shown). System size dependence is similar to the one presented in~\cite{Li08} with the smaller systems developing large finite-size effects for higher $\kappa$. Reasonable extrapolation to the thermodynamic limit was therefore only possible for $\kappa\le0.45$ where we see that the extrapolated topological gap remains finite and relatively robust to the variation of the Landau level mixing strength. 

\section{Energy Gaps}
\label{sec:energy-gaps}

We now turn to the energy gap and show that the gap collapses as $\kappa$ is increased, mirroring the collapse of the
wave function overlap, thereby justifying the claim that the latter signals the onset of a phase transition. 
There are several different energy gaps in a fractional quantum Hall system, with different experimental manifestations.
The simplest gap, which we will simply call the ``energy gap'' is the difference in energy between the two lowest eigenvalues of the 
Hamiltonian, for fixed particle number. This gap must become small (i.e., vanishing in the thermodynamic limit) at a phase transition.
Hence, it is the appropriate quantity to compute when looking for a phase transition.
However, the energy gap may not be relevant to transport experiments, which are
insensitive to the gap to neutral excitations. The transport gap is typically deduced in one of two ways,
which we discuss in Appendix \ref{sec:gap-defs-appendix}.
For reasons that are explained there, we primarily use the so-called `exciton gap' to estimate the transport gap.
As shown in Fig.~\ref{fig-1} in Appendix \ref{sec:gap-defs-appendix},
the various different ways of computing the transport gap are broadly consistent
though there are quantitative differences.
In Appendix \ref{sec:gap-defs-appendix}, we also establish the 
connection to previous important work~\cite{Morf02} that estimated the transport
gap in the spherical geometry with $S=3$ including finite thickness but neglecting Landau level mixing.
  
The dependence of the energy gap on $\kappa$ and $w/\ell_0$ is shown
in Fig. \ref{fig:overlap-and-gap}. The gaps at $S=3$ and $S=-1$ both decrease monotonically with $\kappa$ and 
collapse to zero at approximately the same value of $\kappa$, coinciding with the vanishing of the overlaps ($\kappa\sim 0.7-1.0$, depending 
on the width, with larger widths corresponding to larger critical $\kappa$'s). 
This supports the conclusion that the decrease of the overlap signals the approach to a phase transition, rather than just a failure of the trial wave functions.

Moreover, the energy gap is larger 
at $S=3$ than at $S=-1$ for most Landau level mixing strengths.  If the true ground state of the system
were at $S=3$, then we would expect that, in the thermodynamic limit, there would be no gap at $S=-1$ since this would be a state
with $8$ charge $e/4$ quasiholes, leading to gapless excitations. The reduction of the gap
at $S=-1$ relative to the gap at $S=3$ is consistent with this, but the fact that it is not  zero indicates
that we may not be seeing the asymptotic behaviour of the system.
For instance, while the aPf ground state must have higher energy than the MR
ground state (assuming that the latter is the ground state) by an extensive energy difference,
it may still have lower energy than the MR state with $8$ quasiholes at these system sizes.  There are numerical indications that the 
size of the quasiholes is on the order of many magnetic lengths. Therefore they may strongly overlap at these system sizes,
thereby leading to a finite gap for finite size systems \cite{Storni2011}.  Hence, the extrapolation to the thermodynamic 
limit might be a much more delicate procedure then previously appreciated and, in fact, could point to a potential reason for the 
long-noticed discrepancy between calculated energy gaps and experimentally measured gaps 
\cite{Willett88,Eisentstein88,Eisenstein90,Pan99,Pan01,Eisenstein02,Xia04,Choi08,Pan08,Dean08,Kumar10,Nuebler10,Pan11,Liu11,Nuebler-PRL2012,Liu13,Gamez13,Pan2014,Deng14,Reichel14}.

\begin{figure}[]
\begin{center}
\includegraphics[width=8.5cm,angle=0]{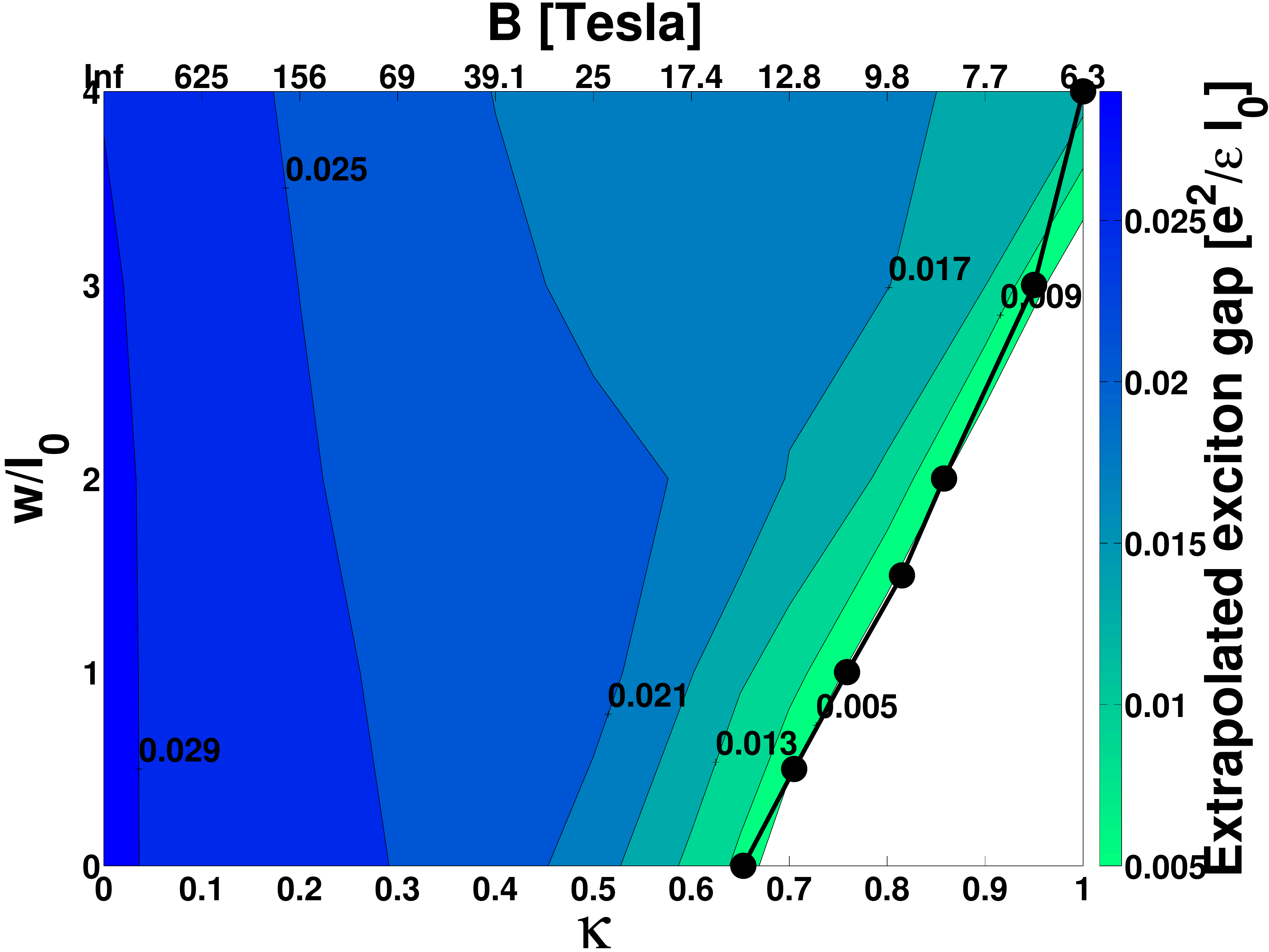}
\caption{(Color online) Color map of extrapolated exciton gap at $S = 3$ versus both well width and $\kappa$.
The contours show specific values of the extrapolated exciton gap.  Note that the extrapolation becomes less reliable when approaching the 
phase transition. The black circles show the transition border determined by the first peak in the entanglement entropy for $N_{\phi}=29$ on 
the sphere (discussed in Sec.~\ref{sec:phase-diagram}). We do not extrapolate the exciton gap after the phase transition (white area). The top $x$-axis is the magnetic field for GaAs samples.}
\label{fig:2Dexciton}
\end{center}
\end{figure}

To provide qualitative guidance to the experiment and to connect to the previous gap estimates in the literature we show in 
Fig.~\ref{fig:2Dexciton} our estimates of the exciton gap extrapolated to the infinite system size. Our results show that Landau level 
mixing and finite-thickness have a non-trivial interplay in the second Landau level.  Landau level mixing reduces the energy gaps more significantly 
than finite-thickness alone.  But we find that both effects, taken together, produce a further reduction.  This is in direct contrast to what has been found in the lowest Landau levels \cite{Bonesteel1995,Scarola2000} where both effects were 
not found to be additive. 

Our results lower energy gaps to bring theoretical estimates even closer to experimental measurements 
\cite{Willett88,Eisentstein88,Eisenstein90,Pan99,Pan01,Eisenstein02,Xia04,Choi08,Pan08,Dean08,Kumar10,Nuebler10,Pan11,Liu11,Nuebler-PRL2012,Liu13,Gamez13,Pan2014,Deng14,Reichel14} 
of the transport gap. Furthermore, the strong suppression of the gap as a function of the Landau level mixing strength that we observe is in good qualitative agreement with the experimental findings presented in Fig. 4 of Ref.~\cite{Samkharadze11}. There, four different experiments are analysed and a similar trend for the dependence of the intrinsic (disorder-corrected) gap on the $\kappa$ parameter is found.

In Appendix~\ref{sec:width-appendix} we demonstrate that width of an infinite quantum well $w$ provides a reasonable parameterization for 
the finite-width effect. In order to compare with experiment one should find the variance of the electron wave function in the direction 
perpendicular to the two-dimensional electron gas in the specific heterostructure (for instance by means of a coupled Schroedinger-Poisson solver in 1D). 
Infinite quantum well width $w$ leading to the same variance should be taken. Note that the width $w$ is given in Fig.~\ref{fig:2Dexciton} 
in units of magnetic length and therefore depends on magnetic field since $\ell_{0}\approx 25 \text{nm}/\sqrt{B[T]}$, where $B[T]$ is the magnetic field in Tesla.

\section{Particle-Hole Symmetry-Breaking Order Parameter}
\label{sec:order-parameter}

States in the MR and aPf universality classes cannot be invariant under
particle-hole symmetry. Indeed, under a particle-hole transformation, a state in the MR universality
class is transformed into a state in the aPf universality class \cite{Lee07,Levin07}.
Thus, if we consider an operator $\phi$ that is odd under a particle-hole transformation, then
$\langle \phi \rangle\equiv\langle \Psi_\mathrm{trial}|\phi|\Psi_\mathrm{trial}\rangle$ must have one sign in any state in the MR universality class
and the opposite sign in any state in the aPf universality class,
assuming that $\langle \phi \rangle$ vanishes only in states that are symmetric under particle-hole
symmetry (i.e., excluding, through a judicious choice of $\phi$,
the possibility that $\langle \phi \rangle$ vanishes `accidentally').
We choose the order parameter to be built from the operator that is conjugate to
the variable $\kappa$ that controls the particle-hole symmetry breaking. This operator is 
$H_\mathrm{3body}=\sum_m V_m^{(3)}(w/\ell_0,\kappa)\sum_{i<j<k}\hat{P}_{ijk}(m)$.  Note that $\kappa$ can be pulled out of this 
expression completely since $H_\mathrm{3body}$ is linear in $\kappa$, hence, we can write $H_\mathrm{3body}=\kappa H^\prime_\mathrm{3body}=\kappa\sum_m V_m^{(3)}(w/\ell_0,1)\sum_{i<j<k}\hat{P}_{ijk}(m)$.  
The order parameter is then taken to be
\begin{equation}
\label{eq:order-parameter-def}
\phi \equiv \frac{1}{2}\left(H^\prime_\mathrm{3body}  - \overline{H^\prime_\mathrm{3body}}\right)
\end{equation}
where the overline denotes particle-hole conjugation.

To demonstrate this definition, let us consider a model that interpolates
adiabatically between the pure Coulomb Hamiltonian and the Hamiltonians whose ground states are in
the MR and aPf universality classes.  That is, $(1-\alpha)H(0,0,1)+\alpha H_3$
or $(1-\alpha)H(0,0,1)+\alpha\overline{H_3}$ where $H_3\equiv \sum_{i<j<k}\hat{P}_{ijk}(m=3)$ is the Hamiltonian that generates 
the MR wave function as an exact zero-energy ground state  and $\overline{H_3}$ is it's
particle-hole conjugate and generates the aPf wave function.  For this model, we take the order parameter
to be  $(H_3 - \overline{H_3})/2$ since $H_3$ is the
operator that breaks the particle hole symmetry by increasing the variable $\alpha$.  The expectation value of this operator has sign
$\langle \Psi_\mathrm{MR} |\phi | \Psi_\mathrm{MR} \rangle <0$ and $\langle \Psi_\mathrm{aPf} |\phi | \Psi_\mathrm{aPf} \rangle > 0$,
and  changes sign in the expected manner, as shown in Appendix \ref{sec:order-parameter-appendix}.  Therefore, we
expect the above definition of $\phi$ (Eq.~(\ref{eq:order-parameter-def}) for the Landau level mixing Hamiltonian) will show similar behaviour and
$\langle \phi\rangle$ will be negative (positive) for an eigenstate in the MR (aPf) universality class.

\begin{figure}[th!]
\begin{center}
\includegraphics[width=8.cm,angle=0]{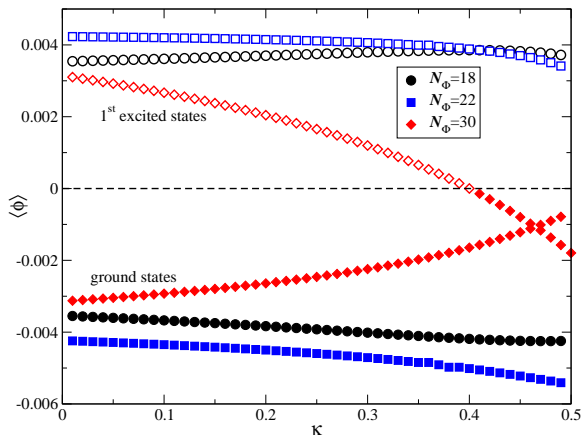}
\caption{(Color online) The expectation value of a particle-hole anti-symmetric order parameter $\phi$ for the 
ground state  and first excited state  of Eq.~(\ref{Heff}) 
on the torus using the hexagonal unit cell for $N_{\Phi}=18$, 22, and 30 as a function of $\kappa$ for $w/\ell_0=0$.
The ground state is consistent with the MR state and has $\langle \phi \rangle <0$ (filled symbols) while the first excited state is consistent with the aPf state 
with $\langle \phi \rangle >0$ (open symbols).  }
\label{fig:order-parameter}
\end{center}
\end{figure}

We first examine this operator in the system in which it is most straightforward.
Recall that on the torus the MR and aPf states occur at the same flux.  
Here $\langle \phi\rangle$ is particularly useful in determining 
the universality class of the ground state of Eq.~(\ref{Heff}).  
The expectation of $\phi$ in the ground state is the most important quantity but we will focus on the lowest and first excited eigenstates on the torus with a hexagonal unit cell containing an odd number of electrons as a
function of $\kappa$ for $w/\ell_0=0$. In Fig.~\ref{fig:order-parameter}, we show
the expectation value of $\phi$ in the ground and first excited states for $N_\Phi=18$, 22, and 30.
These results clearly show that the ground state breaks particle-hole symmetry in the same way as the MR
state and $\langle \phi\rangle<0$. Moreover, the expectation value of $\phi$ in the  first excited states is positive and, therefore,
breaks particle-hole symmetry in the same way as the aPf state.

\begin{figure}[th!]
\begin{center}
\includegraphics[width=8.cm,angle=0]{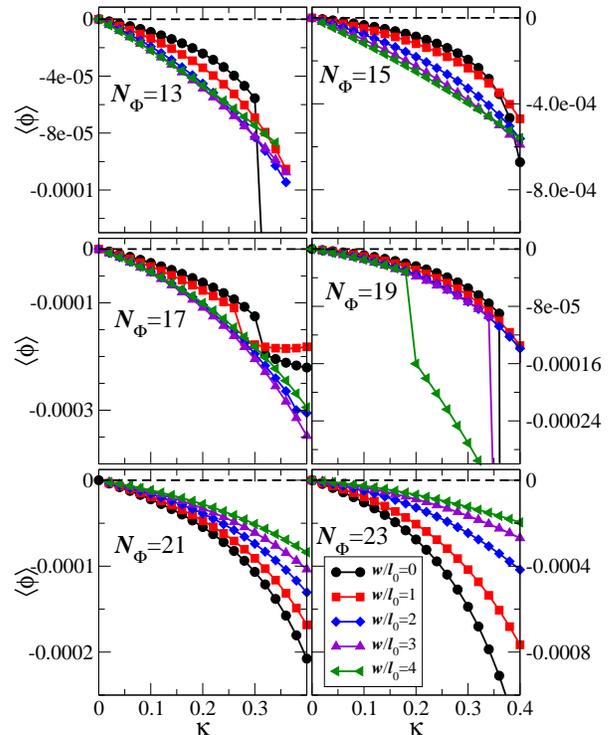}
\caption{(Color online) The expectation value of the particle-hole anti-symmetric order parameter for the ground state of Eq.~(\ref{Heff}) 
in the spherical geometry at the particle-hole symmetric shift $N_{\Phi}=2N_e-1$ for $N_\Phi=13$, 15, 17, 19, 21 and 23 as a function of $\kappa$ for $w/\ell_0 = 0-4$.  Note the $y$-axis is not the same scale for each system size. 
}
\label{fig:sphere-OP-sign}
\end{center}
\end{figure}

\begin{figure}[ht!]
\begin{center}
    \includegraphics[width=8.cm,angle=0]{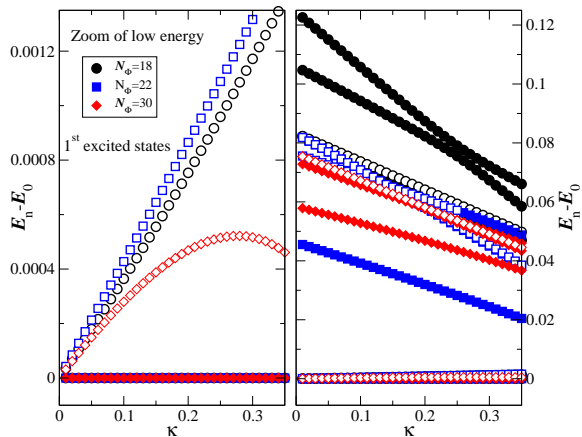}
\caption{(Color online) The total energies relative to the ground state energy  
on the torus using the hexagonal unit cell for $N_{\Phi}=18$, 22, and 30 
as a function of $\kappa$ for $w/\ell_0 =0$.   Similar to Fig.~\ref{fig:order-parameter}, a state with $\langle \phi\rangle < 0$ has a filled symbol while a state  with $\langle \phi\rangle > 0$ 
has an open symbol.  The right panel shows the 6 lowest eigenstates.  We have zoomed in on the lowest two eigenstates in the left panel.   
While many of the higher energy excitations look like they belong to the MR universality class according to $\langle \phi \rangle$, the first 
excited state looks like an aPf ground state rather than an excitation above a MR ground state, unlike the
other low-lying excited states. Note that the energy scale is much larger in the right panel, so the ground and first excited states are
not resolvable there.
}
\label{fig:spectrum-OP-sign}
\end{center}
\end{figure}

In Fig.~\ref{fig:order-parameter} it is observed that $\langle\phi\rangle\neq 0$ for $\kappa=0$.  The hexagonal unit cell has an exact degeneracy for $\kappa=0$ for 
an odd number of electrons in the unit cell as discussed above in Sec.~\ref{sec:overlap}.  At $\kappa=0$ there is a basis in which one of the degenerate states has positive $\langle\phi\rangle$ and the other state has a negative 
value.  This basis evolves smoothly into the $\kappa>0$ eigenstates.  However, we could just as easily take the symmetric and anti-symmetric combinations of these two 
degenerate states, and these combinations would respect the particle-hole symmetry and have vanishing $\langle \phi\rangle$.  For an even number of electrons per unit cell, where 
the degeneracy is not exact at $\kappa=0$, the energy splitting between the symmetric and anti-symmetric combination is non-zero due to tunnelling 
in a finite sized system, and the ground state at $\kappa=0$ is the symmetric combination, with $\langle\phi\rangle=0$. 

Next we consider $\langle \phi\rangle$ in the spherical geometry.  Here we fix $N_\Phi=2N_e-1$ to be the particle-hole symmetric point since the shift $S$ explicitly breaks 
particle-hole symmetry and we want to observe this symmetry breaking due to Landau level mixing effects.  Fig.~\ref{fig:sphere-OP-sign} shows the order parameter for 
the ground state
at $N_\Phi=13$, 15, 17, 19, and 21 for various $w/\ell_0$ as a function of $\kappa$.  
Here the order parameter vanishes for $\kappa=0$ and increasing $\kappa$ drives the system into the
MR universality class and $\langle\phi\rangle$ becomes more negative for increasing $\kappa$.

Finally we investigate the lowest few energy eigenstates of Eq.~(\ref{Heff}) 
in the torus geometry using the hexagonal unit cell for $N_e$ odd in Fig. \ref{fig:spectrum-OP-sign}.  States with negative (positive) order parameter are indicated 
by a filled (open) symbol.  The first excited state has $\langle \phi\rangle > 0$ but the rest have
$\langle \phi\rangle < 0$. Thus, although the second, third, fourth, and fifth excited states look like an exciton on the MR ground state, in that they 
have a negative expectation value of the order parameter and therefore belong in the MR universality class, 
the first excited state does not. It, instead, looks like the aPf state. This is consistent with conclusions from the overlaps,
but can only occur in small systems.

\section{Entanglement Properties and Phase Diagram}
\label{sec:phase-diagram}

From the preceding calculations, we have seen the following concomitant behaviors:  a sharp drop in the energy gap, a corresponding drop in the overlap between the ground state and the MR wave function, 
a negative expectation value of a particle-hole symmetry-breaking order parameter. The first of these
vanishes at the phase transition to a competing phase. 

\begin{figure}[bht!]
\begin{center}
\includegraphics[width=7.5cm,angle=0]{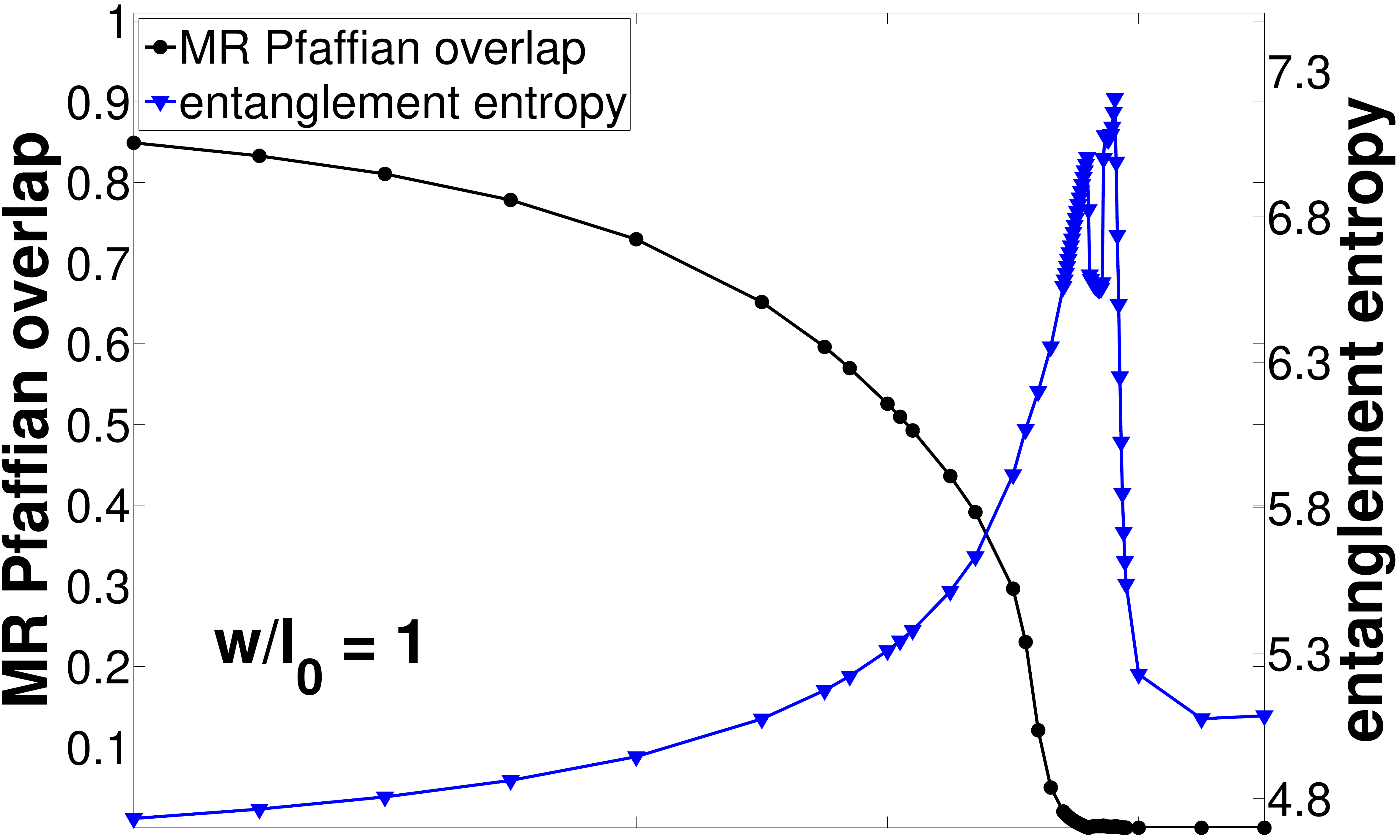}\\
\includegraphics[width=7.5cm,angle=0]{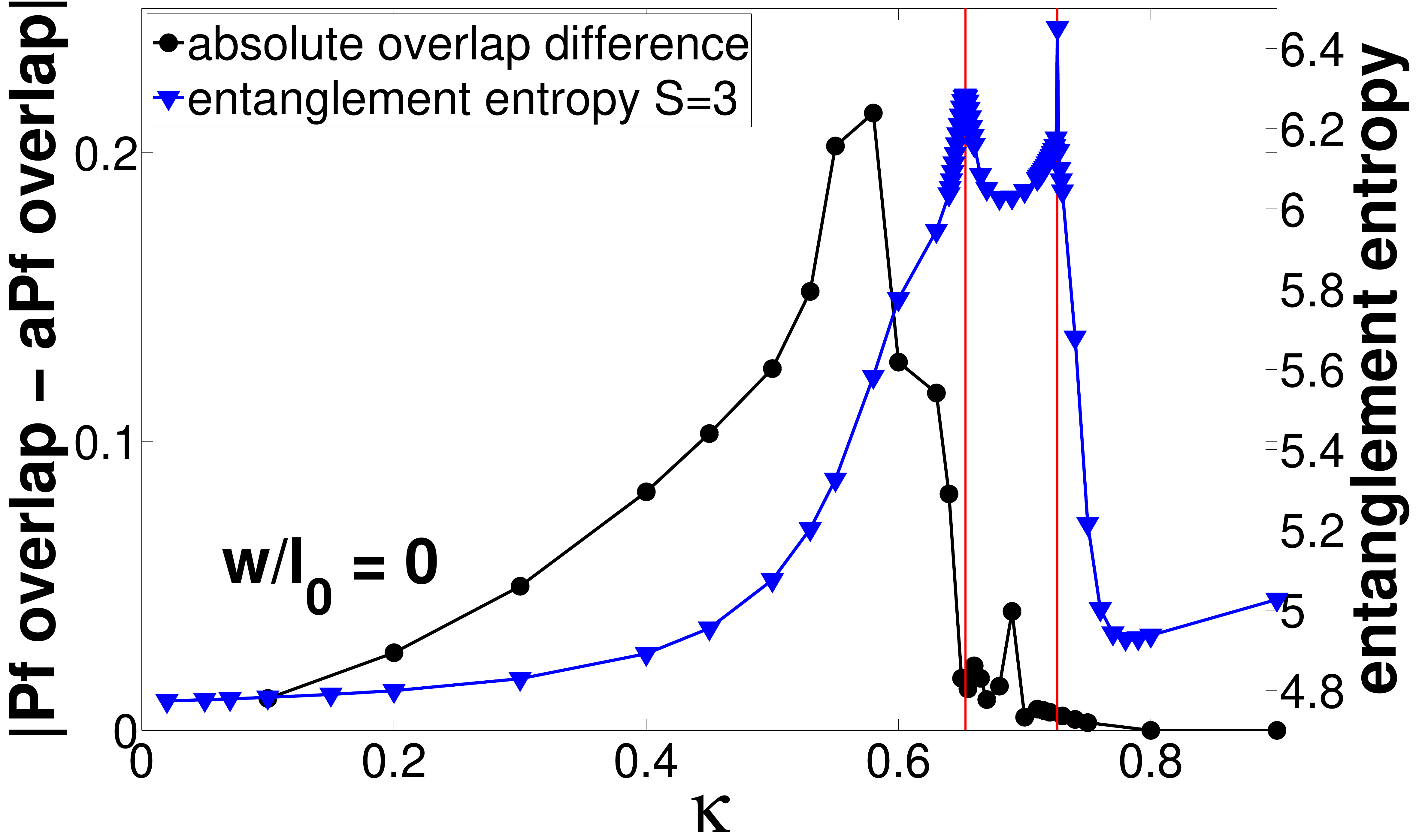}

\caption{(Color online) Top Panel: Entanglement entropy and wavefunction overlap with the MR state  for width $w/\ell_0= 1 $. Bottom panel: Entanglement entropy and the absolute difference of the MR Pfaffian and aPf overlaps for width $w/\ell_0= 0 $. In both cases the spherical geometry was used with $N_{\Phi}=29$ and $S=3$.  The red vertical lines in the bottom panel are just a guide to the eye to indicate the peaks in the entanglement entropy. }
\label{EE-peaks}
\end{center}
\end{figure}

This phase transition point can also be identified by computing
the bipartite entanglement entropy, which is the von Neumann entropy of the reduced density matrix
~\cite{Levin06,Kitaev06b,Haque07,Zozulya07,Biddle11}, discussed in Sec. \ref{sec:ES}.
Fig.~\ref{EE-peaks} shows that the resulting entanglement entropy displays two nearby peaks as a
function of $\kappa$ (only a single peak for $w/\ell_0>1.5 $).  The position of the two peaks coincides
with the vanishing of the overlap which, in turn, coincides with the vanishing of
the energy gap, as per Fig.~\ref{fig:overlap-and-gap}.
These peaks in the entanglement entropy indicate phase transitions~\cite{Amico2008}.

In the bottom panel of Fig.~\ref{EE-peaks}, we see that there are two distinct peaks in the entanglement entropy at $\kappa=0$
at $w/\ell_0=0$, while in the top panel, we see that the two peaks are barely distinguishable at $w/\ell_0=1$.
Intriguingly, the ground state has higher overlap with the aPf wavefunction in the intermediate phase between
the two entanglement entropy peaks although both overlaps are quite small. We speculate about this in Sec.~\ref{sec:conclusions}.

It is instructive to discuss the nature of all the phases mentioned in relation with the corresponding entanglement spectra for $w/\ell_0= 0$ shown
in Fig.~\ref{fig-ES-all}. Comparing the spectra at $\kappa=0.2$ and $\kappa=0.53$ in Fig.~\ref{fig-ES-all} (upper and lower left panels) we observe that, with increasing Landau level mixing strength, the universal part of the entanglement spectrum gets absorbed by the ``generic" spectrum above it. This leads to a decrease of the "topological gap"~\cite{Li08} as shown in Fig.~\ref{fig-top_gap_S3}.

\begin{figure}[]
\begin{center}
\includegraphics[width=9cm,angle=0]{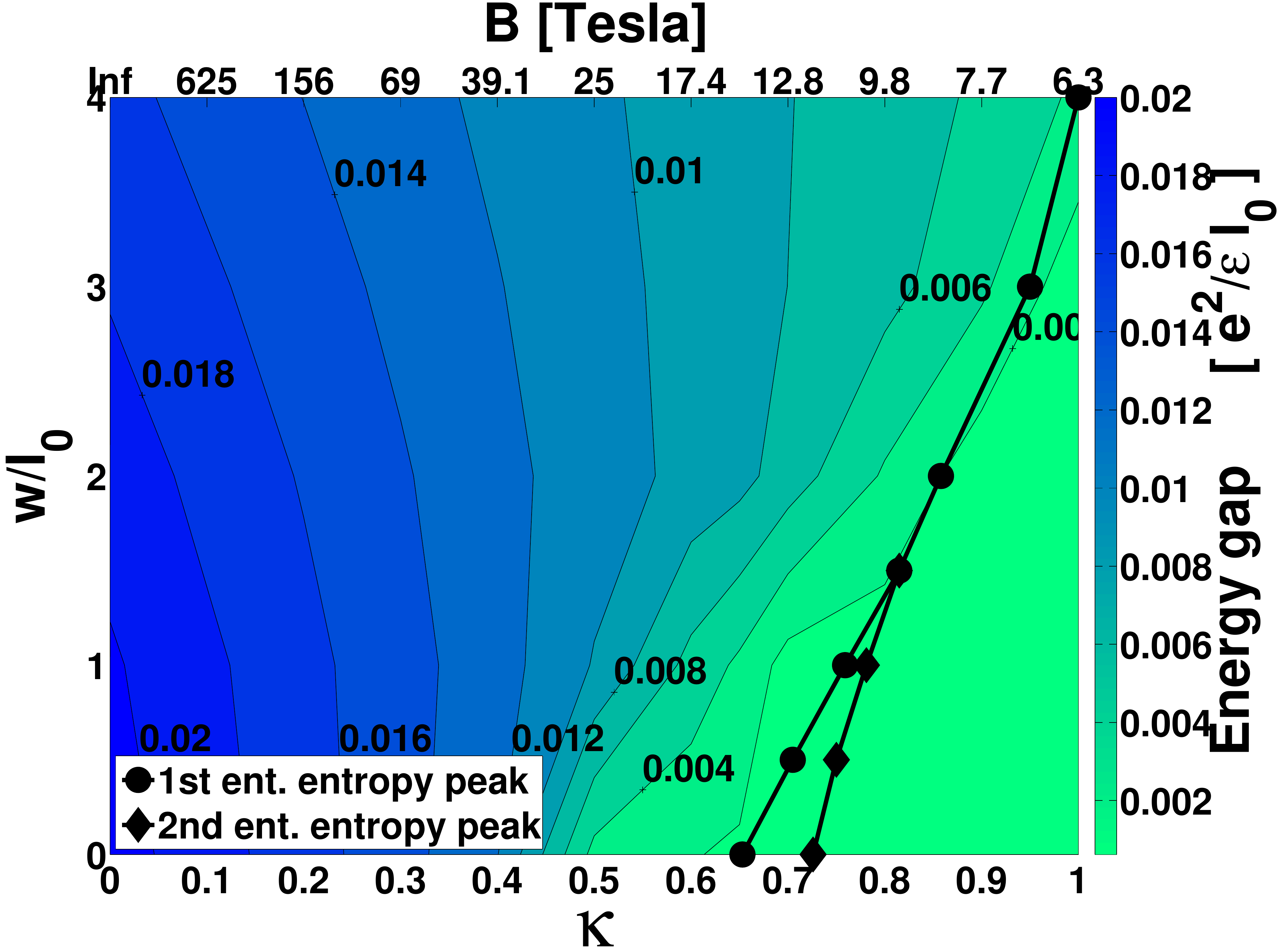}
\caption{(Color online) Quantum phase diagram obtained from a color map of the energy gap (difference between the two lowest eigenstates for the system with $N_{\phi}=33$) plotted versus well width and $\kappa$.  Contours plot specific values of 
the extrapolated gap.  Lines with black circles and black diamonds show the positions of the entanglement entropy peaks for $N_{\phi}=29$ and represent the approximate phase boundaries. All the states to the left from the line connecting the black circles belong to the MR Pfaffian universality class. The top 
$x$-axis is the magnetic field for GaAs samples.  All results are obtained on the sphere.}
\label{fig:phase-diagram}
\end{center}
\end{figure}

Between the two peaks of entanglement entropy, at approximately $\kappa=0.66$, the low-lying levels of the spectrum have positive slope (upper right panel of Fig.~\ref{fig-ES-all}). This indicates that in this phase there exists an edge mode propagating in the direction opposite to the MR Pfaffian edge. However, both the energy gap and entanglement gap are quite small,
so to say anything definitive about this state would require much larger system sizes (compared to what is currently available using exact diagonalization).  In  the phase diagram shown in Fig.~\ref{fig:phase-diagram}, this state is located between the two black lines indicating the entropy peaks.

The entanglement spectrum for stronger Landau level mixing has a completely different nature as shown in the bottom right panel
of Fig.~\ref{fig-ES-all}. We defer the discussion of this regime to later work.

Finally, we discuss an approximate quantum phase diagram (QPD) for the FQHE
at $\nu=5/2$  in Fig.~\ref{fig:phase-diagram}. The 
QPD is determined with two distinct measures: energy gap and entanglement entropy.  The energy gap depicted is
 for the largest system with $N_{\Phi}=33$ while the entanglement entropy is  for $N_{\Phi}=29$.  Fig.~\ref{fig:phase-diagram} shows a contour plot of the energy gap for $S=3$, 
as functions of $\kappa$ and $w/\ell_0$.  We also indicate the position of the first peak in the entanglement entropy (black circles),
clearly showing that it occurs where the energy gap becomes very small for ${N_e}=18$ (presumably indicating that it vanishes
in the thermodynamic limit).  The results presented in Fig.~\ref{fig:phase-diagram} are in agreement with overlaps with the MR state as well.  
This QPD can serve as a guide for experimental searches for robust FQHE at $\nu=5/2$ and is the first approximate QPD calculated at $\nu=5/2$ including both Landau level mixing 
and finite width.

\section{Conclusions}
\label{sec:conclusions}

Our results demonstrate that the $\nu=5/2$ state for small non-zero $\kappa$ and
$0\leq w/\ell_0 \leq 4$ is in the universality class of the Moore-Read Pfaffian state. In the small $\kappa$ limit,
our approximations are controlled: we use the correct Hamiltonian to $O(\kappa)$,
and all corrections to our Hamiltonian are of higher-order in
$\kappa$ and, therefore, can be neglected for sufficiently small $\kappa$.
Our results are in qualitative agreement with the results of
Ref. \onlinecite{Wojs10}. We reached our conclusion by computing several properties of the ground state.  They all 
validate the use of overlaps in this case.
Our results are in disagreement with the results of Ref. \onlinecite{Rezayi09}, which found a
ground state in the aPf universality class. 

Finite size effects might be a potential source of error in our study. Especially, since we are using for our finite-system calculations the pseudopotentials, originally derived for the infinite system. From our available extrapolations (see Appendix~\ref{sec:finiteSize-sphericalpp}) we observe however that the mentioned approximations do not change our results qualitatively.

Finally, the phase that emerges at $\kappa$ just larger than the gap closing
is an interesting open problem. The energy gap and the entanglement gap are too small for us to say anything
reliable at present. However, the overlap with the aPf is larger than the overlap with the MR state
and the entanglement spectrum is consistent with a counter-propagating edge mode, so it is possible that
the aPf state occurs in this narrow window, albeit with much smaller energy gap (possibly more in line with experimental
gap values). Another possibility is a strong pairing phase \cite{Read00}.

Note that small $\kappa$ corresponds to relatively high magnetic fields, e.g., $\kappa=0.5$ is a magnetic field
of $25$ T for GaAs samples. The range of magnetic fields and quantum well widths over which there is a $\nu=5/2$ state in both
experiments and our numerics is the
range $6 \text{T}\leq B \leq 12$ T and $w/\ell_0\sim2-4$. For $B\lesssim 6$ T, we do not find a quantum Hall state at $5/2$ even though
experiments see a $5/2$ plateau all the way down to $B\sim 1-2$ T~\cite{Dean08,Pan2014}.
There are two distinct possible explanations
for this discrepancy between our results and experiments.  One is that our effective Hamiltonian is simply not
quantitatively correct for $\kappa \gtrsim 1$ and including higher order corrections in $\kappa$ would shift the phase transition to lower magnetic fields.  Additionally, for widths beyond $w/\ell_0\gtrsim5$, real experimental systems are often better 
described as two-component systems.  The other possibility is that the experimental observations at fields below $\sim 6$ T are 
spin unpolarized states -- a possibility that we have ignored in this work since we have assumed that
the system is fully spin-polarized.  It is an open question as to the effect of Landau level mixing and finite width have on the spin-polarization 
and whether the ground state, if unpolarized, is or is not in the universality class of the MR Pfaffian or aPf phase.  These questions will have to await future studies.

Although our model does not allow to precisely predict the critical magnetic field corresponding to the vanishing of the FQHE gap our results demonstrating the strong suppression of the gap by Landau level mixing are in good qualitative agreement with the experimental observations~\cite{Samkharadze11}.

\appendix

\section{Models for Non-Zero Width}
\label{sec:width-appendix}

We include non-zero width of the two-dimensional electron system using two approaches.  
In the first approach, we, for $\kappa\neq 0$, assume that the electrons are confined to an infinitely deep square quantum well in the
$z$-direction so that the $z$-dependence of the wave function for the $n^\mathrm{th}$ subband is ${\phi}_n(z)=\sqrt{2/w}\sin((n+1)\pi z/w)$ 
with $w\in[0,d]$ and subband energy $\epsilon_n=(n+1)^2\pi^2\hbar^2/(2m_z w^2)$.  Here $m_z$ is the effective 
electron mass in the quantum well (see Ref.~\onlinecite{Peterson13b} for details).  

In the second approach we choose an alternative Gaussian model to 
demonstrate that the above choice of finite thickness model 
does not change our results qualitatively or quantitatively.   
We fix $\kappa=0$ and take the $z$-dependence to have a Gaussian form
${\phi}(z)=(\sigma^2 2\pi)^{-1/4} e^{-{z^2}/{4\sigma^2}}$ (this wave function is the solution of a parabolic potential but 
since we use it only for $\kappa=0$ we do not consider any subband mixing effects). 

\begin{figure}[h]
\begin{center}
\includegraphics[width=8cm,angle=0]{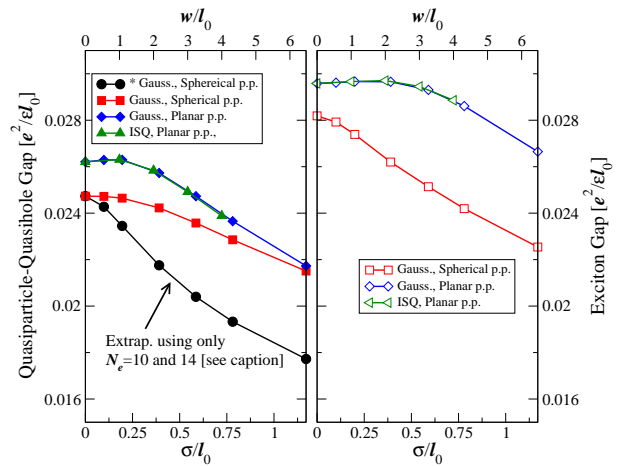}
\caption{(Color online) Extrapolated quasiparticle-quasihole (left panel, filled symbols) and exciton (right panel, open symbols) energy gaps as a function of well width.  Each gap is 
calculated with spherical and planar pseudopotentials (p.p.)  for two different models of the finite thickness, Gaussian and infinite square well (ISQ), cf. Appendix~\ref{sec:width-appendix}.  
The solid black circles (marked in the legend with an asterisk $^*$) are the same as the solid red squares but are obtained from an extrapolation of only $N_e=10$ and 14 electrons. 
The solid black circles reproduce the width dependence obtained in Ref.~\onlinecite{Morf02} (Fig.~14b)  to show that using more particles in the extrapolation (the other data points 
use $N_e=10,14$ and 18 electrons) leads to larger gaps. } 
\label{fig-1}
\end{center}
\end{figure}

Fig.~\ref{fig-1} shows that the
energy gaps (extrapolated to the thermodynamic limit) are very similar for both models of non-zero width.  
To compare each model at a similar width we considered each energy gap as a function of the variance of the wave functions, 
$\mathrm{var}=\sqrt{\langle z^2\rangle - \langle z \rangle^2}$, that is, $\mathrm{var}=\sigma/\ell_0$ for the Gaussian wave function 
and $\mathrm{var}=0.180756(w/\ell_0)$ for the infinite square well wave function.

\section{Exciton and Quasiparticle-Quasihole Gaps}
\label{sec:gap-defs-appendix}

To estimate the gap on the sphere in the thermodynamic limit,
we generally follow Ref.~\onlinecite{Morf02}.  We take energy differences and perform a linear (in $1/N_{e}$) extrapolation to infinite system sizes after we multiply the energy difference by the factor $\sqrt{N_\Phi/2N_e}$, \cite{Morf02}.   
We calculate energy difference (and therefore the gap) in two distinct ways:

\noindent
(i) The exciton gap is the energy difference between the ground and the 
lowest excited state with total angular momentum $L=N_e/2$ or $L=N_e/2-1$ for $N_e/2$ even or odd, respectively. This excited state contains a quasi-particle and  
quasi-hole, maximally separated on the sphere~\cite{Morf02}.   The quasi-particle and quasi-hole are assumed to have charges $\mp e/4$
and to be separated by the diameter of the sphere $2\sqrt{N_e}\ell_0$ so we subtract the energy of the quasi-particle-quasi-hole ideal Coulomb attraction 
$-\frac{1}{32}\frac{1}{\sqrt{N_e}}$ (this is $A_{q=1/4}(\nu=1/2)$ in Ref.~\onlinecite{Morf02}). This exciton gap is calculated for $N_e=8, 10, 14, 16, 18$ ($N_e=12$ is 
aliased with a composite fermion state at $\nu=3/5$ and is ambiguous~\cite{jain2007composite}). Note that the background energy does not enter into this definition of the gap since its contribution explicitly cancels.

\noindent
(ii) Alternatively, one can compare the ground state energy at $N_\Phi$ to the ground state energies with one additional/fewer flux quantum.  These
are states with two quasiholes or two quasiparticles, respectively.  
While more subtle than in the case of the exciton gap,  the background energies cancel again. The resulting gap, sometimes called the quasiparticle-quasihole (qp-qh) 
gap ~\cite{Morf02}, is calculated for $N_e$=10,14, and 18.  Other system sizes are aliased (see Table \uppercase\expandafter{\romannumeral3\relax} in Ref.~\onlinecite{Morf02}).

Although we present both gap calculations, we consider the exciton gap a more reliable gap 
estimate in the thermodynamic limit due to the less severe aliasing problem. Only estimates using exciton gap are used in the main text.

In Fig.~\ref{fig-1} we illustrate the differences between the various ways of calculating the thermodynamic limit 
of the energy gap. The gap is roughly the same for an infinite square well potential as it is for a Gaussian $z$-dependence (once they are taken such that the wavefunction variance is the same)
and for spherical and planar pseudopotentials. However, there are some quantitative differences:  
(i) if one extrapolates based on larger system sizes,  
the width dependence of the qp-qh gap is less pronounced and can only account for a $13\%$ decrease of
the gap compared to $28\%$; (ii)  although in agreement qualitatively, using planar pseudopotentials 
instead of spherical ones tends to give higher gap estimates; and (iii) the exciton gap 
is larger than the qp-qh gap.

\section{Particle-Hole Symmetry Breaking Order Parameter for an Illustrative Model Hamiltonian}
\label{sec:order-parameter-appendix}

\begin{figure}[h!]
\begin{center}
\includegraphics[width=8.25cm,angle=0]{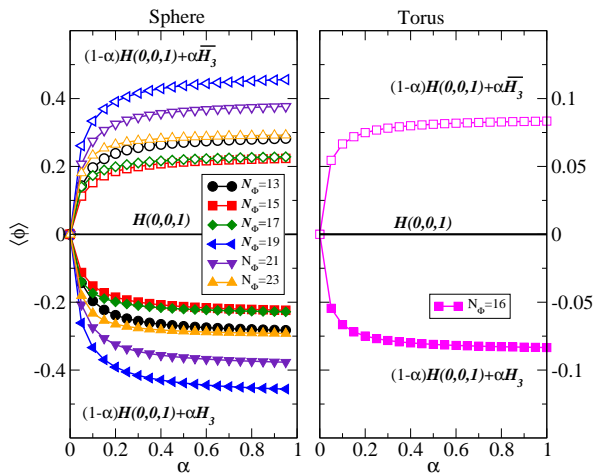}
\caption{(Left) The particle-hole anti-symmetric order parameter computed for Hamiltonians in the spherical geometry that interpolate between the $N=1$ Landau level 
Coulomb Hamiltonian $H(0,0,1)$ at $\alpha=0$ and the Hamiltonians
$H_3$ (solid symbols) and $\overline{H_3}$ (open symbols), respectively, for systems with $N_\Phi=13-21$ corresponding to $N_e=7-11$.  (Right) The same but 
computed on the torus using a rectangular unit cell for $N_e=8$.
}
\label{fig:OP-illustrative}
\end{center}
\end{figure}

In Section \ref{sec:order-parameter}, we introduced an order parameter (Eq.~(\ref{eq:order-parameter-def}))
for particle-hole symmetry breaking. We now show that this order parameter has negative expectation
value in the MR trial wave function and positive expectation value in
the aPf trial wave function.  Fig.~\ref{fig:OP-illustrative}
shows $\langle \phi \rangle$ for the ground state of $(1-\alpha) H(0,0,1)+ \alpha {H_3}$ and the ground state
of $(1-\alpha) H(0,0,1) + \alpha \overline{H_3}$. Here ${H_3}\equiv \sum_{i<j<k}\hat{P}_{ijk}(m=3)$ and $\overline{H_3}$ is the particle-hole
conjugate of $H_3$. As may be seen from the $\alpha\rightarrow 1$ behaviour in Fig.~\ref{fig:OP-illustrative}, when the ground state is in the MR 
universality class, $\langle\phi\rangle <0$ and when it is the aPf wave function, $\langle\phi\rangle >0$. Moreover, the order
parameter interpolates smoothly between zero and these values, as $\alpha$ is increased from zero. 

These calculations were performed on the sphere and on the torus for comparison.  For the spherical geometry, $\langle \phi \rangle$ is 
calculated
at $N_\Phi = 2N_e - 1$, which is the particle-hole symmetric value of the shift on the sphere. Thus, at $\alpha=0$, there is no particle-hole
symmetry-breaking due to finite-size effects. At $N_\Phi=2N_e-1$ the ground state of ${H_3}$ is not the
MR wave function, but the MR wave function with 4 MR quasiholes, and the ground state of $\overline{H_3}$ is the aPf wave function with
4 aPf quasiparticles.  On the torus, we use the rectangular unit cell and show the results for the corner of the Brillouin zone for $N_e=8$ electrons, i.e, 
$\mathbf{K}=(N_0/2,N_0/2)$.  The other 
$\mathrm{K}$ points corresponding to the MR state display similar behaviour.  Note in this choice of unit cell we find $\langle\phi\rangle=0$ for $\alpha=0$, 
while in the hexagonal unit cell for an
even number of electrons this is not the case.   

\section{Finite-Size Effects, Planar and Spherical Pseudopotentials}
\label{sec:finiteSize-sphericalpp}

In our model Hamiltonian, we used planar pseudopotentials.
Although spherical pseudopotentials approach planar ones in sufficiently large systems,
our use of planar pseudopotentials can be a source of systematic error in small spherical systems.
In this section, we analyze the differences between spherical pseudopotentials and the planar
ones used in the results reported in section~\ref{sec:qualitative}. We also perform an
extrapolation in system size to ensure our conclusions hold in the thermodynamic limit.
We restrict our discussion here to small $\kappa$, which is the limit in which our Hamiltonian
is exact on the plane.

\begin{figure}[h!]
\begin{center}
\includegraphics[width=8.25cm,angle=0]{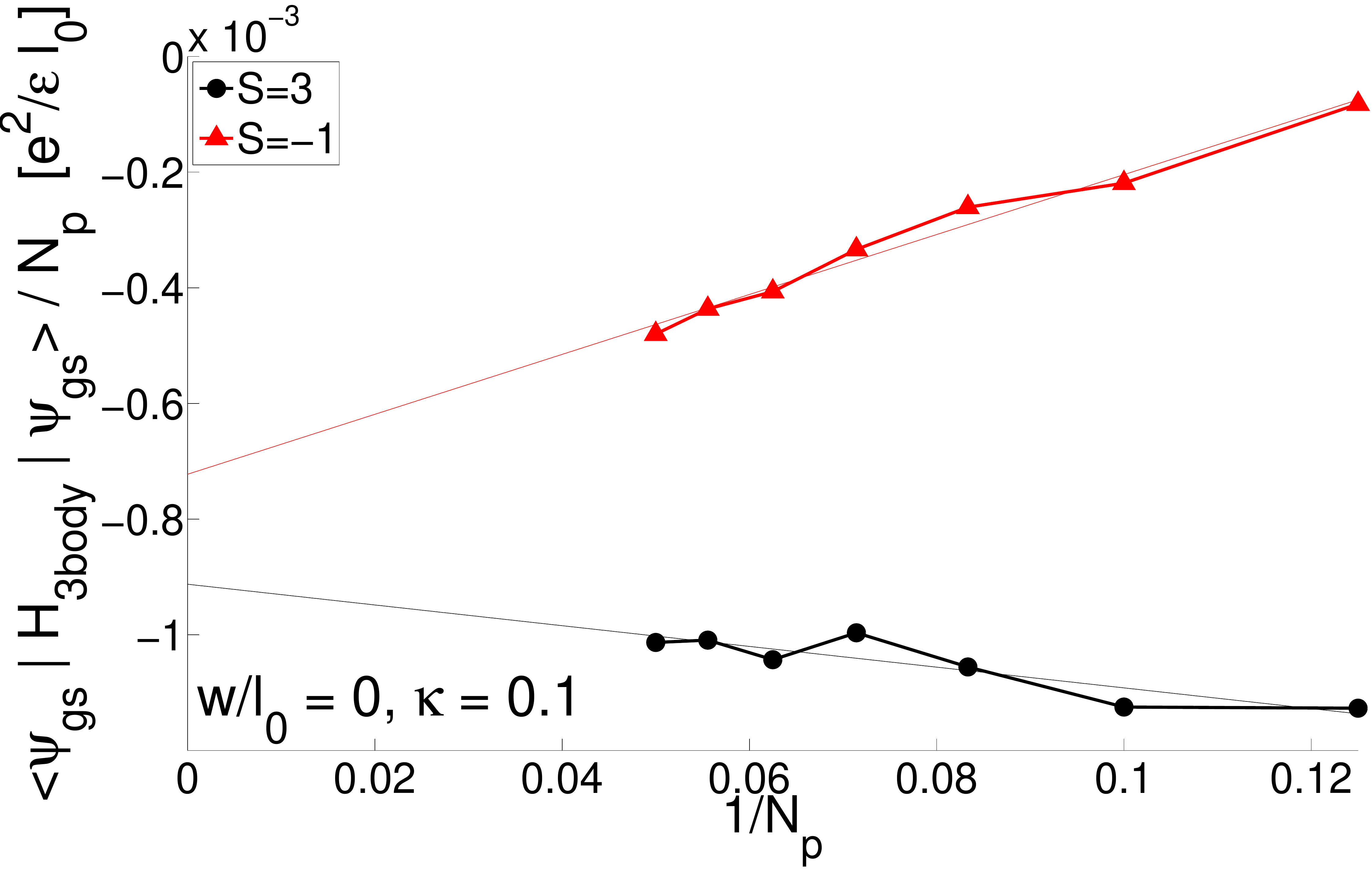}\\
\includegraphics[width=8.25cm,angle=0]{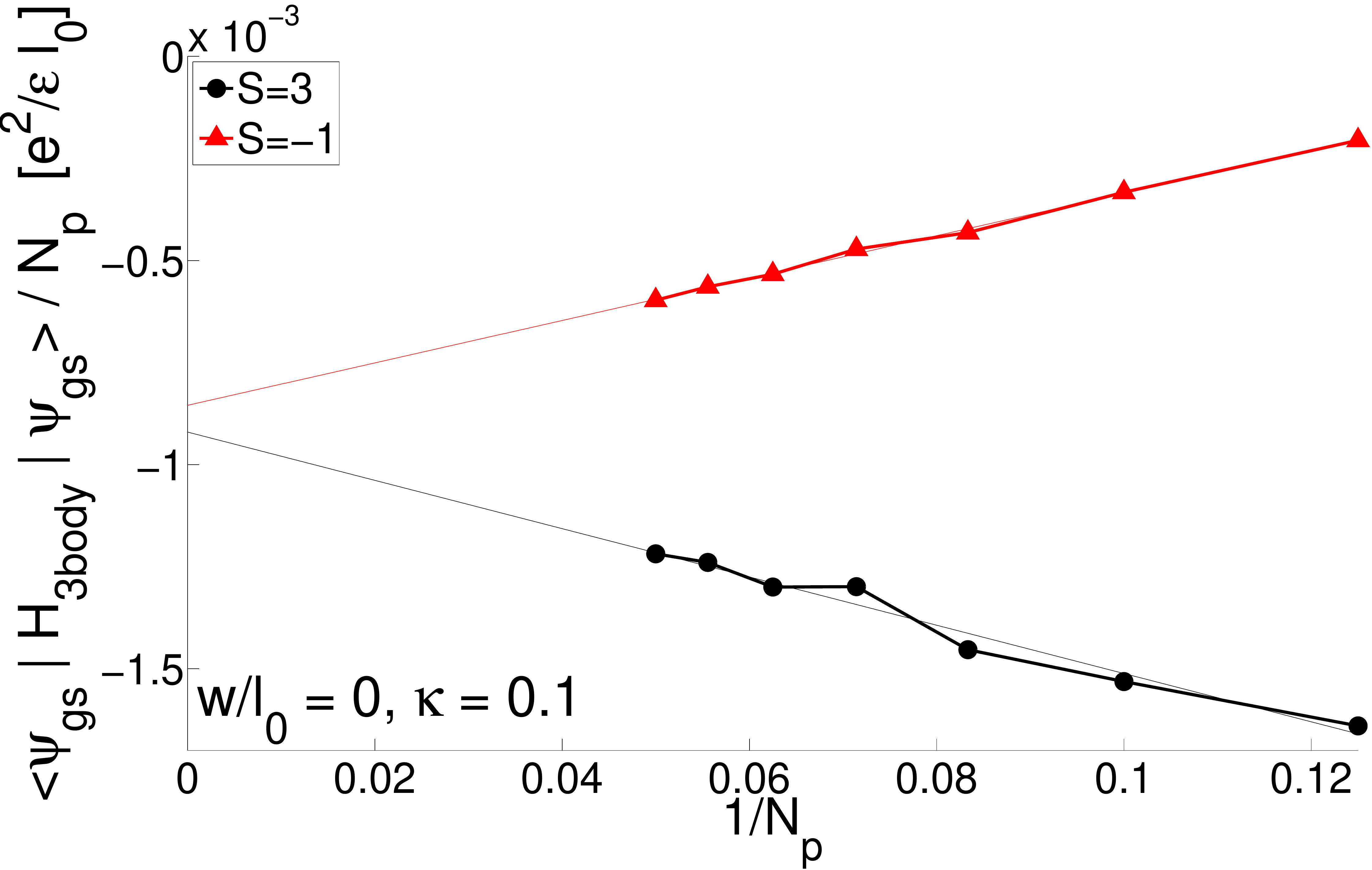}
\caption{ Three-body energy contributions are calculated with system size-dependent (bottom panel) and extrapolated (top panel) spherical pseudopotentials. 
The energy difference between the extrapolated values are 0.000066 $e^2/\epsilon\ell_0$ = 0.045 $\kappa |V^{(3)}_3|$ (bottom panel) and 0.00019 $e^2/\epsilon\ell_0$ = 0.13 $\kappa |V^{(3)}_3|$ (top panel). Although relatively small, these values are per particle and should thus provide extensive energy separation between the two states in the thermodynamic limit. 
}
\label{fig:afs_avgH3_spherical_pp}
\end{center}
\end{figure}

To consider the effect of planar versus spherical pseudopotentials in Eq.~(\ref{Heff}), we
compute the spherical pseudopotentials using a program kindly provided by Steve Simon, which was also used in Ref.~\cite{Rezayi13}. We obtain the spherical three-body pseudopotentials for each of our relevant system sizes along with the three-body pseudopotentials carefully extrapolated to infinite size. For our biggest systems the pseudopotentials could not be calculated directly and we used the values obtained from extrapolation in $1/N_{\Phi}$. The pseudopotentials dependence on $1/N_{\Phi}$ we find is somewhat softer than presented in~\cite{Rezayi13} but has a clear linear dependence hence using the extrapolated values is justified. The code used~\cite{Rezayi13} only gives us the differences of the pseudopotentials (e.g., $V_5-V_3$). A constant shift of the three-body pseudopotentials does not influence the many-body state. We choose this shift so that the finite-size spherical $V^{(3)}_3$ is equal to the
extrapolated planar $V^{(3)}_3$.

In Fig.~\ref{fig:afs_avgH3_spherical_pp}, we display the lowest-order perturbative (per particle)
energy contributions of the three-body terms of $H(w/\ell_0=0, \kappa=0.1, 1)$ given in Eq.~(\ref{Heff})
using spherical pseudopotentials, rather than the planar pseudopotentials used in Fig.~\ref{fig:afs_avgH3_extrapolated}.
The top panel of the Fig.~\ref{fig:afs_avgH3_spherical_pp} uses the pseudopotentials obtained by extrapolating the spherical pseudopotentials to the thermodynamic limit. In principle, this should be precisely the same as in
Fig.~\ref{fig:afs_avgH3_extrapolated}, but there are small differences since the extrapolation from these system sizes
does not give precisely the planar values. The lower panel, in turn, shows the same expectation values
with spherical pseudopotentials used at each system size. Three-body contributions were again evaluated in the Coulomb ground state in both cases. The results are qualitatively consistent with those obtained using planar pseudopotentials:
the energy is lowered more at $S=3$ than at $S=-1$ for each individual system size as well as in the thermodynamic limit.

\acknowledgments
C.N. and M.R.P. have been supported by the DARPA QuEST program. C.N. has been supported by the
AFOSR under grant FA9550-10-1-0524. M.R.P. thanks the Office of Research and
Sponsored Programs at California State University Long Beach. M.T. and K.P. were supported by the Swiss National Science 
Foundation through the National Competence Center in Research QSIT and by the European Research Council through ERC Advanced Grant SIMCOFE.
K.P., M.T., and M.R.P. are grateful to Microsoft Station Q for its hospitality during part of the completion of this work.
C.N. and M.T. acknowledge the hospitality of the Aspen Center for Physics, supported by NSF grant 1066293.  V.W.S. acknowledges support from AFOSR under grant FA9550-11-1-0313. 
Th. J. acknowledges computer time allocation CNRS-IDRIS-100383. This work was supported by a grant from the Swiss National Supercomputing Centre (CSCS) under project s395.
K.P. is grateful to R. Morf, A. Wojs and S. Simon for many helpful discussions.
Th. J. acknowledges discussions with I. Sodemann, A. H. MacDonald and J. K. Jain.
C.N. thanks R. Mong and M. Zaletel for discussions. We thank S.Simon for supplying us with the code to calculate system-size dependent spherical pseudopotentials.

\bibliography{diag5half}

\end{document}